\documentclass[%
 reprint,
 amsmath,amssymb,
 prd,
]{revtex4-2}

\usepackage{graphicx}% Include figure files
\usepackage{makecell}
\usepackage{dcolumn}% Align table columns on decimal point
\usepackage{bm}% bold math
\usepackage[export]{adjustbox} % Para ajustar la posición de la figura
\usepackage{array}
\usepackage[caption=false]{subfig}
\usepackage[usenames,dvipsnames,svgnames,table]{xcolor}
\usepackage{hyperref}% add hypertext capabilities
\definecolor{verdeoscuro}{rgb}{0, 0.5, 0}

\begin{document}

\preprint{APS/123-QED}

\title{High-frequency gravitational waves detection \\ with the BabyIAXO haloscopes}% Force line breaks with \\

\author{José Reina-Valero,$^{a*}$ Jose R. Navarro-Madrid,$^b$ Diego Blas,$^{c,d}$ \\ Alejandro D\'iaz-Morcillo,$^b$ Igor Garc\'ia Irastorza,$^e$ Benito Gimeno,$^a$ Juan Monz\'o-Cabrera,$^b$ }
\affiliation{$^a$Instituto de Física Corpuscular (IFIC), CSIC-University of Valencia, Calle Catedr\'atico Jos\'e Beltr\'an Martínez, 2, 46980 Paterna (Valencia), Spain, \\ 
$^b$Departamento de Tecnolog\'ias de la Informaci\'on y las Comunicaciones,
Universidad Polit\'ecnica de Cartagena,
Plaza del Hospital 1, 30302 Cartagena, Spain,\\
$^c$Institut de F\'isica d’Altes Energies (IFAE), The Barcelona Institute of Science and Technology, Campus UAB, 08193 Bellaterra (Barcelona), Spain\\
%Grup de Física Teòrica, Departament de Física, Universitat Autònoma de Barcelona,
%08193 Bellaterra (Barcelona), Spain \\
$^d$Instituci\'o Catalana de Recerca i Estudis Avan\c cats (ICREA), Passeig Llu\'is Companys 23, 08010 Barcelona, Spain \\
$^e$Center for Astroparticles and High Energy Physics (CAPA), Universidad
de Zaragoza, Zaragoza 50009, Spain}

% E-mail addresses: only for the corresponding author
\email{jose.reina@uv.es}

\date{\today}% It is always \today, today,

\begin{abstract}
We present the first analysis using RADES-BabyIAXO cavities as detectors of high-frequency gravitational waves (HFGWs). In particular, we discuss two configurations for distinct frequency ranges of HFGWs: Cavity 1, mostly sensitive at a frequency range of 252.8 - 333.2 MHz, and Cavity 2, at 2.504 - 3.402 GHz, which is a scaled down version of Cavity 1. We find that Cavity 1 will reach sensitivity to strains of the HFGWs of order $h_1\sim 10^{-21}$, while Cavity 2 will reach $h_2\sim 10^{-20}$. These represent the best estimations of the RADES-BabyIAXO cavities as HFGWs detectors, showing how this set-up can produce groundbreaking results in axion physics and HFGWs simultaneously.
\end{abstract}

\maketitle

%\tableofcontents

%%%%%%%%%%%%%%%%%%%%%%%%%%%%%%%%%%%%%%%%%%%%%%%%%%%%%%%%%%%%
\section{\label{sec:intro}Introduction}
%%%%%%%%%%%%%%%%%%%%%%%%%%%%%%%%%%%%%%%%%%%%%%%%%%%%%%%%%%%%

After their first direct discovery in 2015 by the LIGO collaboration \cite{LIGOScientific:2016aoc} in the 100 Hz band, gravitational waves (GWs) are reshaping our understanding of the cosmos. More concretely,  these messengers bring new information from previously undetectable events, and may be the key to access the most primordial instances of the Universe. When one considers the extraordinary progress in cosmology and astrophysics when the whole spectrum of electromagnetic (EM) radiation was used to explore the Universe, it is hard to think what may happen when GWs in a broad spectrum of frequencies will be discovered. 

Out of the different possibilities, the `high frequency' band of GWs has recently received increasing attention~\cite{Aggarwal:2020olq}. Different reasons explain the current interest: first, the signals above the band of groundbase detectors (few kHz) are not highly populated by known astrophysical events, leaving a rich landscape of possible signals beyond the standard model to search for without foregrounds; 
second, higher frequency means smaller and faster devices, more compatible with laboratory scales, where one can leverage the revolution in sensing to look for these feeble signals.
%and the currently built laser interferometers based on Fabry-Perot cavities lose sensitivity above few kHz. The reasons for the current excitement about pushing this boundary is coming from a twist of the previous two points: first, the lack of standard signals leave a rich landscape of possible signals beyond the standard model to search for; second, by compressing the size of the detectors, 

From the universal coupling of gravitation, the possibilities to search for HFGWs in the sensing frontier are multi-fold. In this work, we will focus on one of the most promising ones: their impact on microwave electromagnetic cavities immersed in an intense magnetostatic field. 
A GW propagating along an intense static magnetic field can convert into an electromagnetic wave of the same frequency \cite{gertsenshtein1962wave,gertsenshtein_1962}. If this happens inside a cavity of the right size, this signal will accumulate, allowing the eventual detection of the primordial signal. This simple idea has been recently discussed in detail in \cite{Berlin:2021txa,Berlin:2023grv,Ratzinger:2024spd}. The final sensitivity depends not only on the ability to detect tiny EM fields, but also on properties of the cavity, such as its volume or quality  factor. 
%It is hence natural to explore which is the optimal search strategy to look for GWs from the most advanced microwave electromagnetic cavities used to search for axions (and hence immersed in intense magnetic fields). 

In this work, we will focus on  the BabyIAXO helioscope, and derive results for GWs of frequencies between 252.8 to 333.2 MHz and 2.504 to 3.402  GHz.  A novel approach to the problem will be performed since no system with tuning capability and ports for both axions and GWs has been studied yet. We shall discuss how to adapt the %BabyIAXO 
experimental setup to detect HFGWs, and perfom a quantitative study on how well the available system couples to an incident HFGW. This study, involving a tuning system, is specially relevant for the detection of stationary sources, following the same philosophy as in the axion case. Hence, superradiant black-holes and GWs backgrounds are of special interest for this kind of setup~\cite{Berlin:2021txa}.

This paper is organized as follows: firstly, a brief commentary about the BabyIAXO experiment is performed, explaining its setup and capabilities; secondly, the electromagnetic design of the cavity for the RADES-BabyIAXO haloscope as well as tuning system and ports is showed, discussing its geometry and advantages. After this, a numerical calculation of the form factor between the HFGW and RADES-BabyIAXO cavities is made paying special attention to the several values that experimental parameters can adopt. In addition, a calculation of the sensitivity estimation is performed, and the expression for the scanning rate is obtained with a formulation similar to the dark matter axion case.

%%%%%%%%%%%%%%%%%%%%%%%%%%%%%%%%%%%%%%%%%%%%%%%%%%%%%%%%%%%%
\section{The BabyIAXO Experiment}\label{sec:babyiaxo}
%%%%%%%%%%%%%%%%%%%%%%%%%%%%%%%%%%%%%%%%%%%%%%%%%%%%%%%%%%%%

The IAXO (International Axion Observatory) experiment is one of the most ambitious projects in axion searches, whose aim is to detect solar axions \cite{IAXO2014,IAXO2019}. The intermediate stage, the BabyIAXO helioscope, will be built as a stepping stone for the experiment, and it will be composed of a cryostat and a dipole magnet with two bores of 0.7 metres in diameter and 10 metres in length, providing a transversal magnetostatic field of 2 T \cite{BabyIAXO2021}. This magnet gives the opportunity, in addition to operating as a helioscope, to search for dark matter axions with haloscopes, and the RADES collaboration made a proposal to use resonant cavities at low frequencies to search for dark matter axions in the mass range of 1-2 $\mu$eV, expecting competitive values in terms of axion sensitivity \cite{BABYIAXO_RADES}. To cover this mass range, four cavities were designed with a frequency tuning system that allows to shift the operational frequency during the data taking campaigns. Although these four cavities share the same shape, the main difference lies in the diameter, which is larger for the cavity operating at lower frequencies and smaller for the cavity operating at higher frequencies, all maintaining the same length of 5 metres. Therefore, two cavities can be placed in each bore during the measurement operation, giving the chance to search in parallel at different masses (using four different cavities), or enhance the sensitivity by summing coherently the signals (using four cavities with the same dimensions). These cavities will be set in a cryostat system based on cooling via cryocoolers and a closed helium circulation loop~\cite{cryostat}, in order to cool them down to  4.6 K. Plans to push this temperature down to tens of mK are set as part of the recently granted ERC-SyG \textit{DarkQuantum} with dilution refrigerator (DL) technology. This will allow to effectively get noise figures close to the standard quantum limit with superconducting quantum interference devices (SQUIDs). 

\begin{figure}[ht]
    \centering
    \includegraphics[scale=0.4]{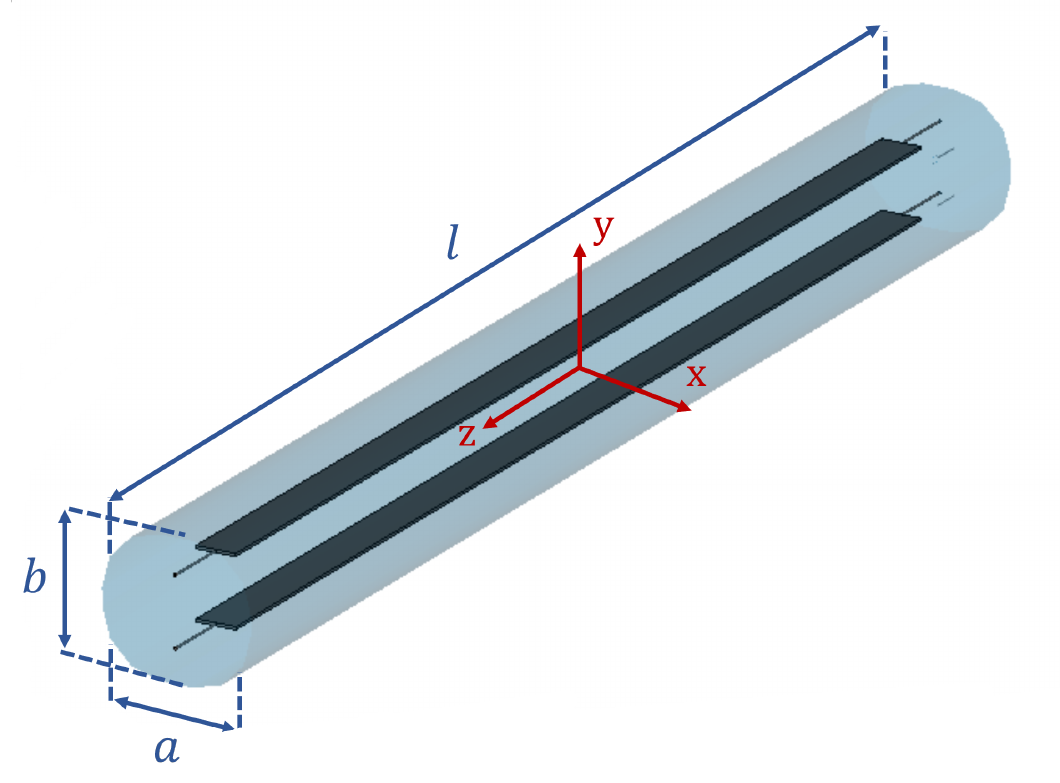}
  \caption{
 Schematic of the RADES-BabyIAXO cavities geometry. The cavity (in blue color) has a quasi-cylindrical shape, with dimensions $a$ (width), $b$ (height) and $l$ (length), where the relation $b = 0.9a$ is fulfilled. The metallic rectangular plates for the frequency tuning are shown in dark grey color. In red, the chosen reference frame in the center of the cavity.
    }
    \label{fig:cav_babyiaxo}
\end{figure}

As mentioned above, the properties of the RADES-BabyIAXO haloscope allow for the search of HFGWs, which can be done during the axion measurement time by making small changes in the cavities designed for dark matter search.  A novelty of RADES-BabyIAXO  is its kind of tuning mechanism, which comes with its own challenges. Indeed, the introduction of this mechanism perturbs considerably the electromagnetic modal field structure inside the cavity and the coupling between the HFGW and the resonant system is strongly modified. As a result, the tuning system originally employed for scanning over a mass range of axions can be useful for a frequency sweep to detect HFGWs with different frequencies.
%As reported in \cite{BABYIAXO_RADES}, a calculation of the sensitivity was performed obtaining $h_{0}$ $\sim$ $10^{-21}$. However, it should be noted that this(2.5 - 3.4) estimation did not take into account the tuning plates, .  

%%%%%%%%%%%%%%%%%%%%%%%%%%%%%%%%%%%%%%%%%%%%%%%%%%%%%%%%%%%%
\section{Electromagnetic design}\label{sec:emdesign}
%%%%%%%%%%%%%%%%%%%%%%%%%%%%%%%%%%%%%%%%%%%%%%%%%%%%%%%%%%%%

The electromagnetic study of the RADES-BabyIAXO cavities was performed in \cite{BABYIAXO_RADES}, where a quasi-cylindrical shape was adopted in order to avoid the degeneration of the modes $\mathrm{TE_{111}}$ and maximize the volume, but not to fill completely the bore in order to leave some space for cables and instrumentation at both sides. Two cavities are analysed in this article: the first one is the lower-frequency cavity that has been designed to operate in the MHz range (VHF and UHF bands), called Cavity 1, and the second one is a 10 times scaled-down version of the first cavity working at the GHz range (S-band), called Cavity 2. This second cavity would not be used in RADES-BabyIAXO, but has been designed and manufactured as a test model of Cavity 1 for the GHz range. The tuning system incorporated in both cavities consist of two metallic thin rectangular plates along the longitudinal axis that turn simultaneously, shifting the frequency of the operational EM mode. A schematic of the cavity geometry is shown in Fig. \ref{fig:cav_babyiaxo}, while Table \ref{tab:Cav_dimensions} contains the dimensions of the two designs.

\begin{table}[ht]
\centering
\begin{tabular}{| c | c | c  | c | c |}
\hline
 Cav. model & a (cm) & b (cm) & l (cm) & Freq. range  \\
\hline
Cavity 1 & 56 & 50.4 & 500 & 252.8 - 333.2 MHz  \\ \hline
Cavity 2 & 5.5 & 4.9 & 50 & 2.504 - 3.402 GHz \\

\hline
\end{tabular}
\caption{Dimensions and operational frequency range of the studied cavities. The thickness of both metallic plates is 1 cm.}
\label{tab:Cav_dimensions}
\end{table}

It is well known that in dark matter axion experiments using a cylindrical cavity immersed within a dipole magnet, the resonant mode that provides the maximum coupling between the radio-frequency (RF) electric field and the external static magnetic field is the $\textrm{TE}_{111}$ mode, which is a degenerate mode that, at the same resonant frequency, has two polarizations orthogonal to each other \cite{Pozar}. In the case of the RADES-BabyIAXO cavities, due to the presence of the plates and with the condition $a \neq b$, such degeneracy disappears and both modes, quasi-$\textrm{TE}_{111}$ with vertical and horizontal polarization, adopt two distinct frequencies.

\begin{figure}[ht]
    \centering
    \includegraphics[scale=0.25]{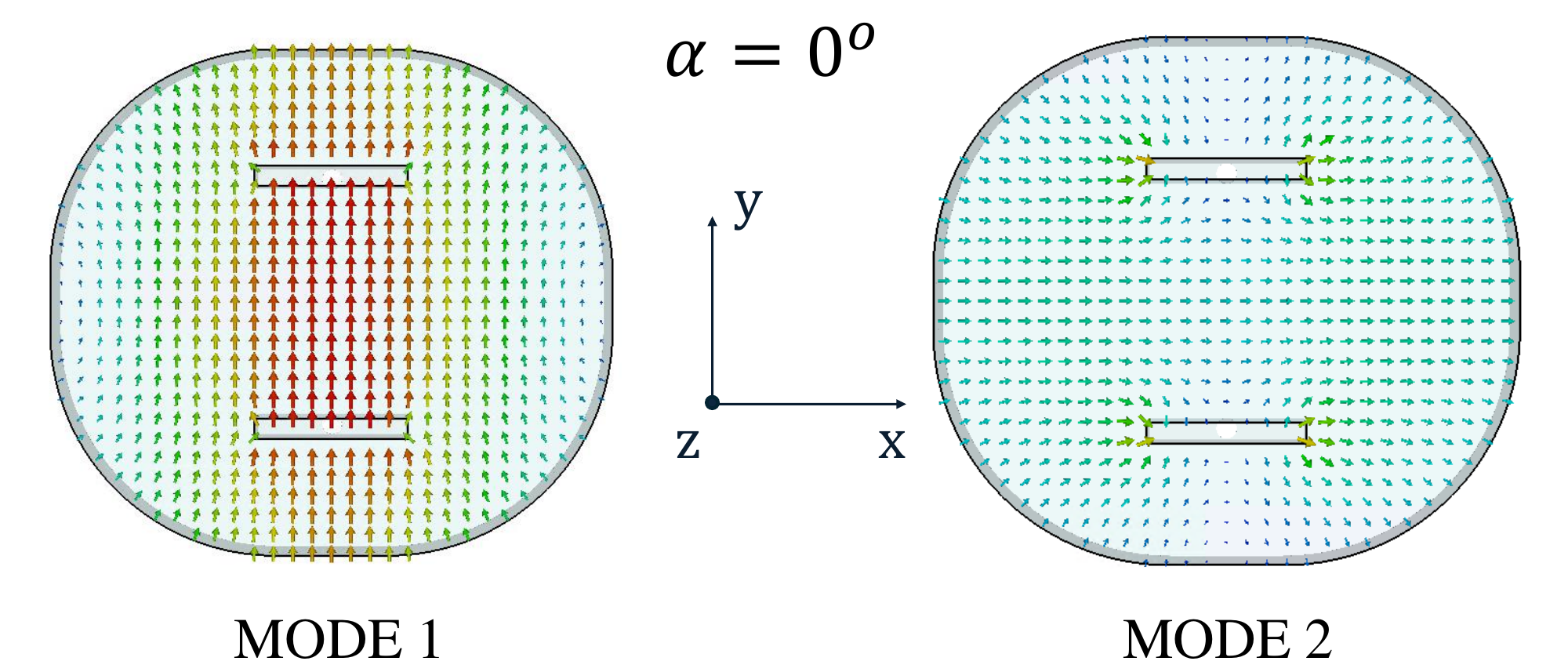}
    \caption{Electric field profiles of Mode 1 and Mode 2 in a transversal cut of Cavity 2 (also applicable to Cavity 1) for plates angle $\alpha$ $=$ $0^{\circ}$. Size and color of the arrows indicate the magnitude of the electric field in linear (from more intense in red to weaker in blue).}
    \label{fig:E_field_profile}
\end{figure}

In HFGWs detection, the use of a second mode is of special interest, as the gravitational wave can couple not only to both modes \cite{Berlin:2021txa,HFGW_pablo}, but it can excite them at the same time if the bandwidth of the HFGW is higher than the frequency gap between both modes. Thus, the modes analysed in this article are the quasi-$\textrm{TE}_{111}$ with an initial vertical polarization (oriented to $y$-axis in Fig. \ref{fig:cav_babyiaxo}) and the quasi-$\textrm{TE}_{111}$ with an initial horizontal polarization (oriented in $x$-axis), which we call Mode 1 and Mode 2, respectively; the RF electric field pattern in the cavity central plane $z=0$ is plotted in Fig. \ref{fig:E_field_profile} for both modes. 
\begin{figure}[ht]
\centering
\subfloat[\label{A}]{\includegraphics[width=0.24\textwidth]{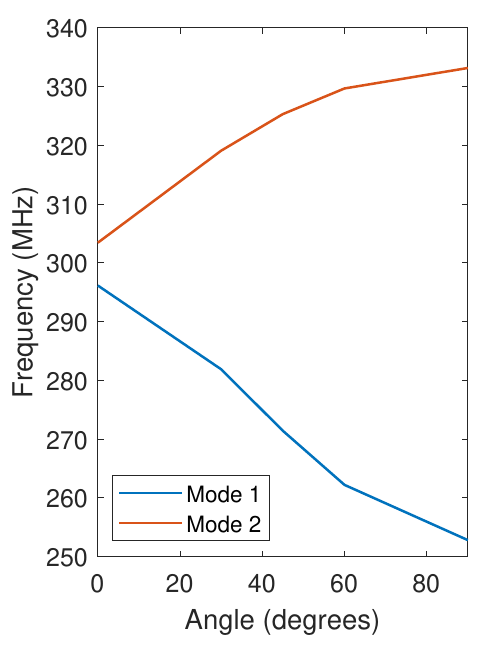}}%
%\hspace{0.1cm} %reemplazar por \hfill si necesario
\hfill
\subfloat[\label{B}]{\includegraphics[width=0.24\textwidth]{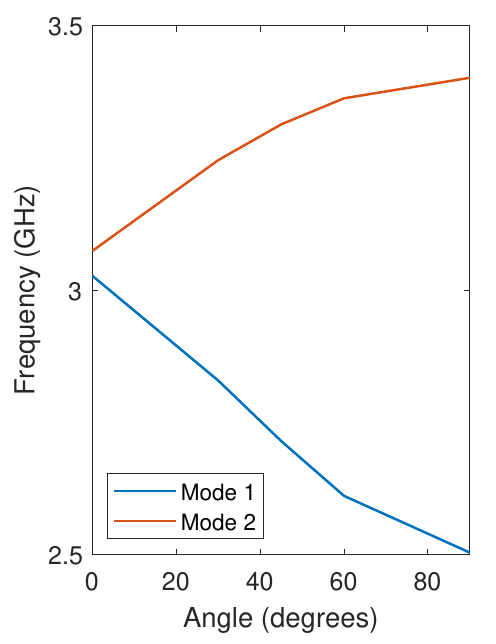}}%
\caption{Resonant frequency variation during tuning versus plates angles for (a) Cavity 1 and (b) Cavity 2.}
\label{fig:freq_tuning}
\end{figure}
These modes are affected differently by the tuning system, as it is shown in Fig. \ref{fig:freq_tuning}.  Mode 1 decreases its resonant frequency when the plates are rotated, while Mode 2 increases its resonant frequency. Note that the tuning range is smaller for Mode 2 than for Mode 1, being this range for Mode 1, 43.33 MHz and 523.5 MHz for cavities 1 and 2, respectively, compared to the range of Mode 2, 29.74 MHz and 327.5 MHz for cavities 1 and 2, respectively.

The possibility to change the resonant frequency of both modes gives the opportunity to search for HFGWs in a total range of 73.1 MHz and 850.9 MHz for Cavity 1 and 2, respectively. However, there is a frequency range where it is not possible to scan, and it corresponds to the frequency separation of both modes at the starting point, i.e. when the angle of the plates is 0$^{\circ}$. To avoid this frequency gap, the cavity section should have a $b/a$ ratio around $b/a \approx 1$.

%%%%%%%%%%%%%%%%%%%%%%%%%%%%%%%%%%%%%%%%%%%%%%%%%%%%%%%%%%%%
\subsection{Coupling ports}
%%%%%%%%%%%%%%%%%%%%%%%%%%%%%%%%%%%%%%%%%%%%%%%%%%%%%%%%%%%%

\begin{figure}[ht]
    \centering
    \includegraphics[scale=0.24]{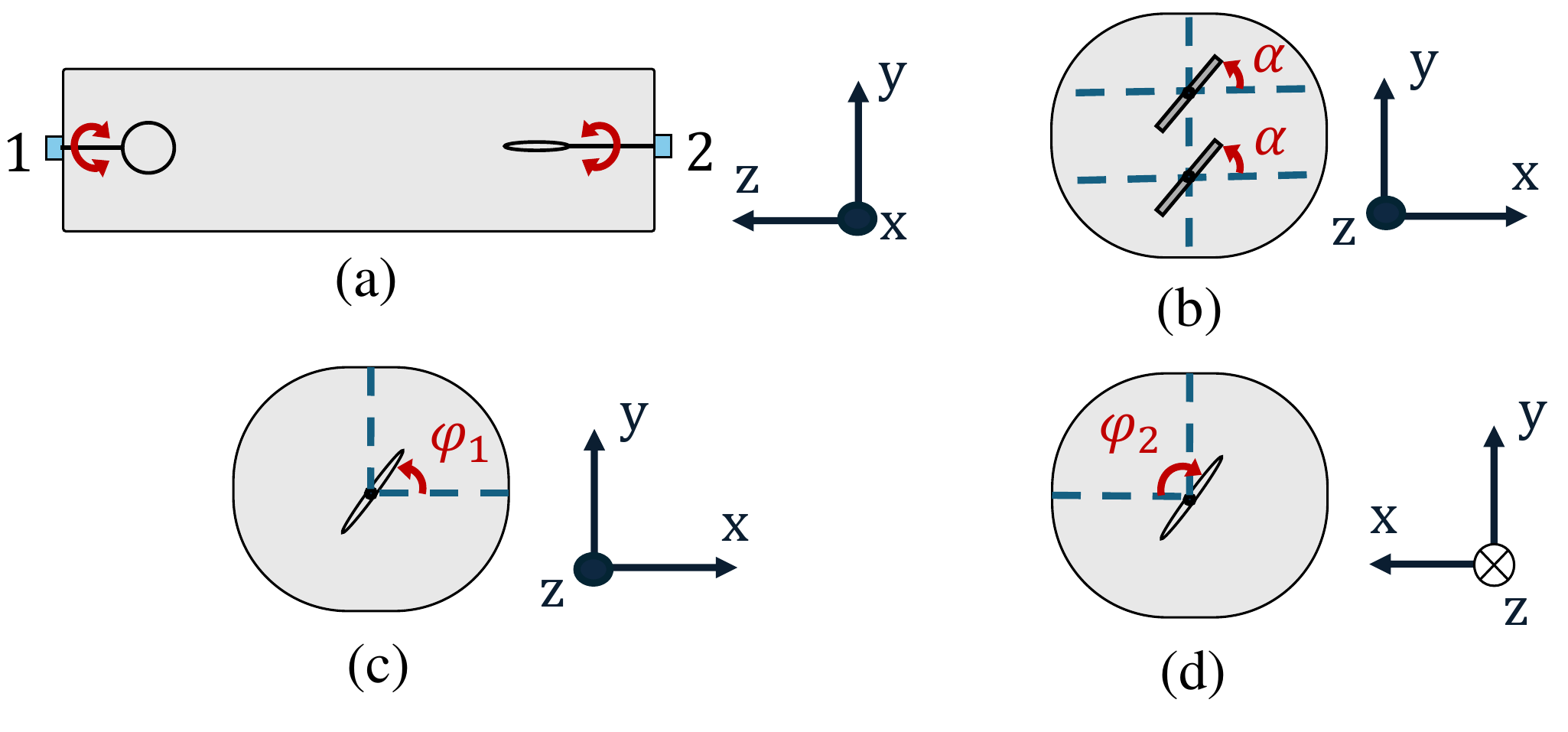}
    \caption{Geometry of ports and plates inside the cavity. (a) Motion representation of the loops for RF magnetic field detection. (b) Definition of angle $\alpha$ of the plates with respect to the $x$-axis. (c), (d) Definition of angles $\varphi_{1}$ and $\varphi_{2}$ with respect to the $x$-axis.}
    \label{fig:ports}
\end{figure}

An important aspect of this experiment is how to couple the EM signals generated by the HFGW, as the position and type of port will determine which modes can be detected. The port, proposed in \cite{BABYIAXO_RADES}, is a coaxial loop located at one of the cavity ends that couples the magnetic field of the Mode 1. In order to couple the Mode 2, the same concept can be used although in this case the second coaxial loop is placed at the other end of the cavity and must be rotated 90$^{\circ}$ with respect to that of the Port 1 in order to couple to the Mode 2 RF magnetic field. Figure \ref{fig:ports} shows a schematic view of the location of the two ports in the haloscope, where Port 1 is a coaxial loop oriented $\phi_1 = 90^{\circ}$ with respect to the \textit{XZ} plane that couples Mode 1. Port 2 is another coaxial loop oriented $\phi_2 = 0^{\circ}$ with respect to the \textit{XZ} plane that couples Mode 2. Rotation angles are denoted as $\varphi_1$ and $\varphi_2$ for the Ports 1 and 2, respectively. Both loops can be rotated in order to re-couple during the frequency tuning process, that is, the coupling coefficient $\beta$ \cite{Pozar} can be varied by rotating the loops. 

%The optimum value for detecting monochromatic HFGWs is $\beta = 4.7$ (critical coupling), which must be re-achieved during tuning, since varying the frequency of the modes will also vary the coupling.

%%%%%%%%%%%%%%%%%%%%%%%%%%%%%%%%%%%%%%%%%%%%%%%%%%%%%%%%%%%%
\section{Form factor}\label{sec:formfactor}
%%%%%%%%%%%%%%%%%%%%%%%%%%%%%%%%%%%%%%%%%%%%%%%%%%%%%%%%%%%%

% Forcing the table to be at the top of the page

This work is focused on the HFGW coupling to the cavity related to the Gertsenshtein effect \cite{gertsenshtein1962wave,gertsenshtein_1962}: GWs will convert into EM waves of the same frequency in the presence of a background very intense magnetostatic field. This coupling can be quantified by the following coupling factor given by \cite{Berlin:2021txa}:
\begin{equation}
    \tilde{\eta}_{m_{+,\times}} = \frac{\left|\int_{V}\vec{E}_{m}\left(\vec{r}\right)\cdot \vec{J}_{+,\times}\left(\vec{r}\right)\mathrm{d}V\right|}{V^{1/2}\left|\int_{V}\vec{E}_{m}\left(\vec{r}\right)\cdot \vec{E}_{m}\left(\vec{r}\right)\mathrm{d}V\right|^{1/2}},
\end{equation}
where $\vec{E}_{m}$ is the RF electric field of the $m$-th resonant mode inside the cavity, $\vec{J_{+,\times}}$ the electric density current arisen by a monochromatic HFGW for plus and cross polarizations (see \cite{Berlin:2021txa}), and $V$ the volume of the cavity.

\begin{figure}[ht]
    \centering
    \includegraphics[scale=0.26]{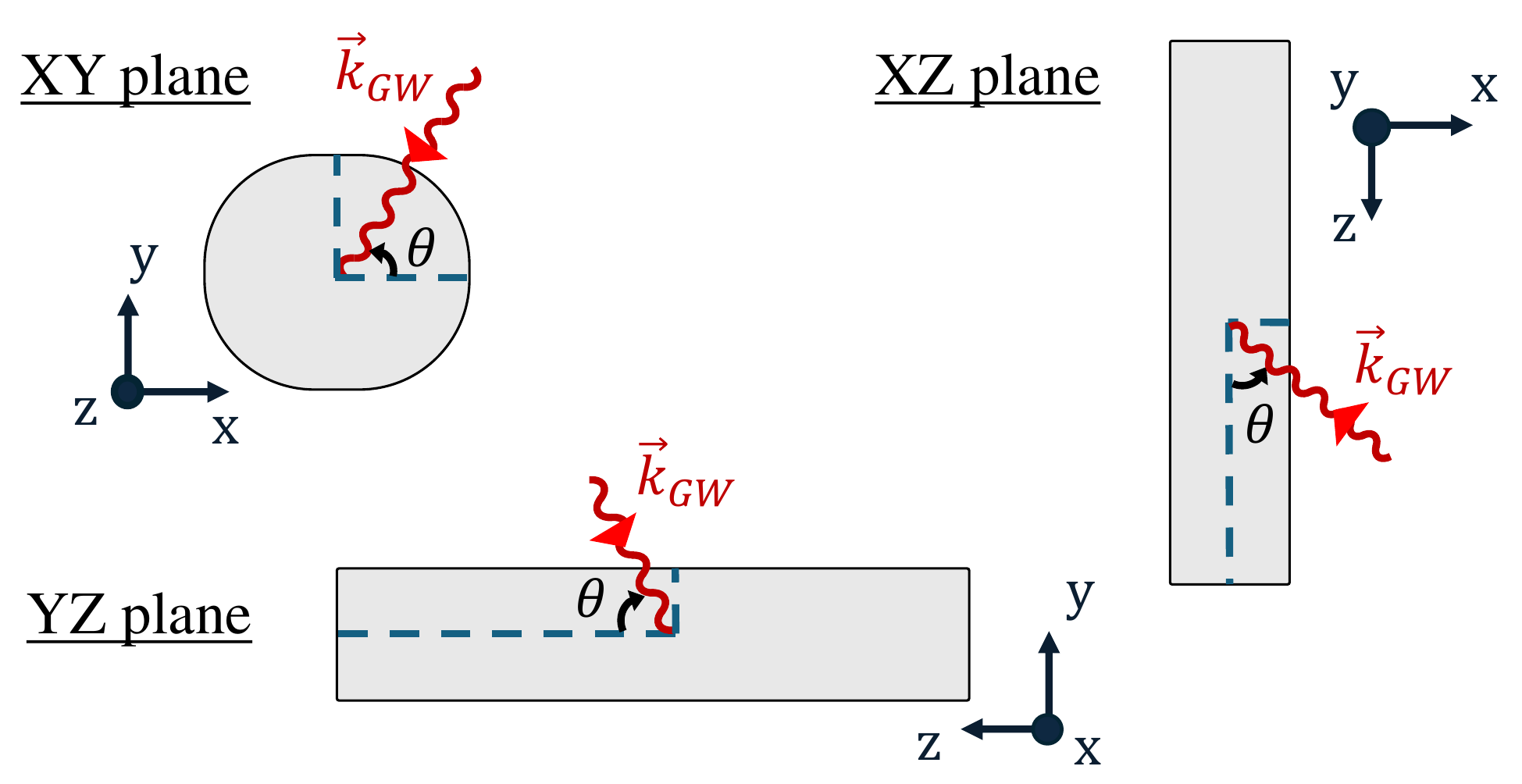}
    \caption{Scheme of HFGWs incidence in each plane.}
    \label{fig:GW_incidence}
\end{figure}

{
\begin{table*}[ht]
    %\captionsetup{type=figure}
    \centering
    \begin{tabular}{| c | >{\centering\arraybackslash}m{0.3\textwidth} | >{\centering\arraybackslash}m{0.3\textwidth} | >{\centering\arraybackslash}m{0.3\textwidth} |}
\cline { 2 - 4 } \multicolumn{1}{c|}{} & \vspace{0.2cm} \hspace{0.12cm} \Large \textit{XY} plane \vspace{0.1cm} & \vspace{0.2cm} \hspace{0.12cm} \Large \textit{XZ} plane \vspace{0.1cm} & \vspace{0.2cm} \hspace{0.12cm} \Large \textit{YZ} plane \vspace{0.1cm} \\
\hline \Large $\tilde{\eta}_{\times}$ & \parbox[c]{\linewidth}{\centering \vspace{.2cm}\includegraphics[scale=0.4]{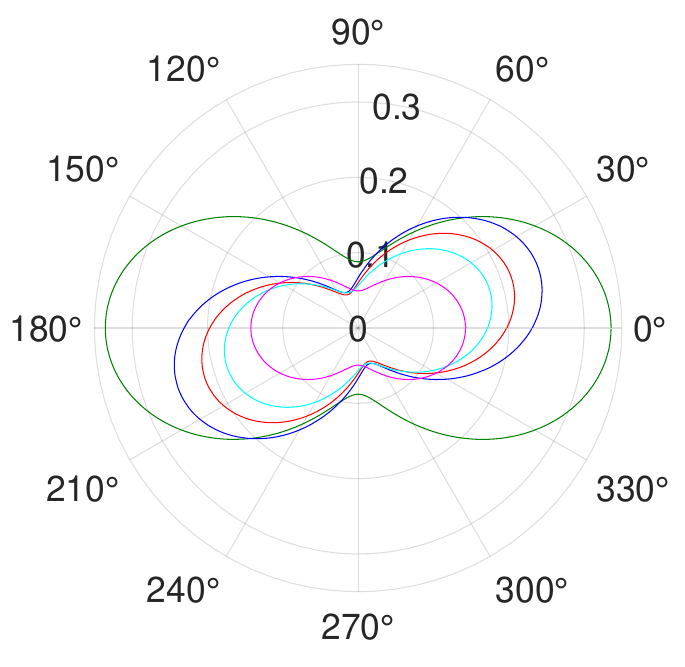}\vspace{.2cm}} & \parbox[c]{\linewidth}{\centering \vspace{.2cm}\includegraphics[scale=0.4]{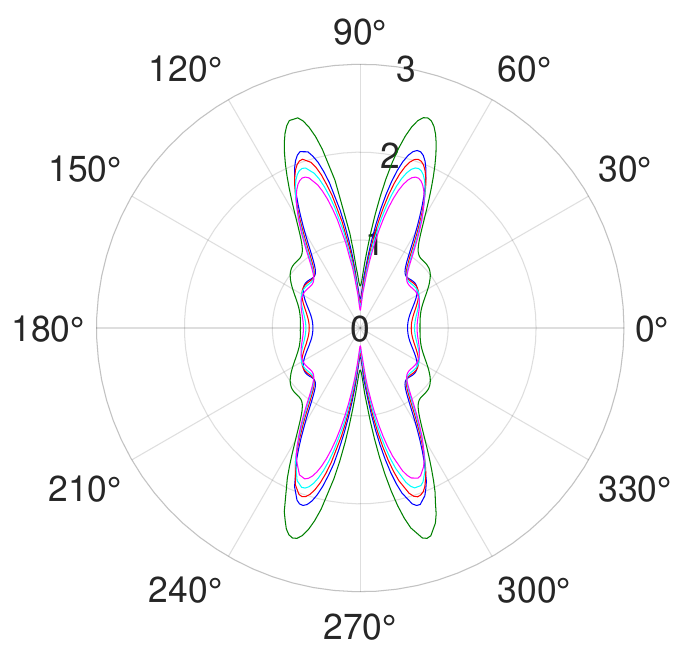}\vspace{.2cm}} & \parbox[c]{\linewidth}{\centering \vspace{.2cm}\includegraphics[scale=0.4]{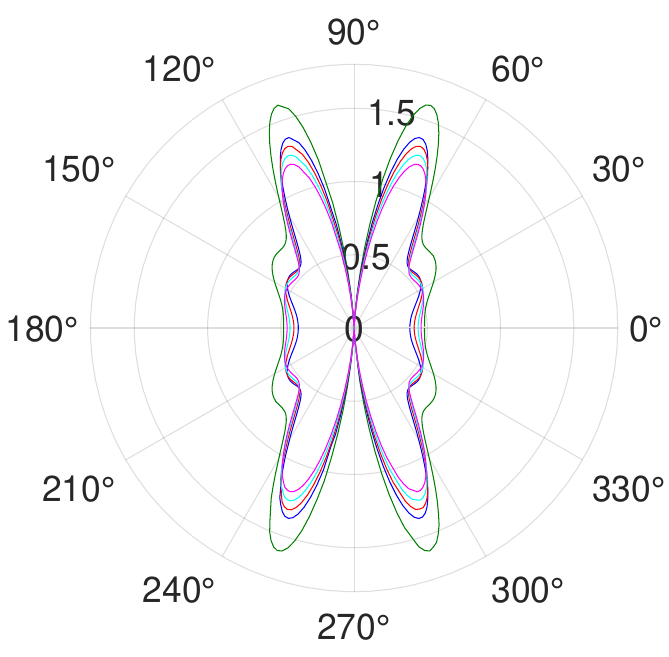}\vspace{.2cm}} \\
\hline \Large $\tilde{\eta}_{+}$ & \parbox[c]{\linewidth}{\centering \vspace{.2cm}\includegraphics[scale=0.4]{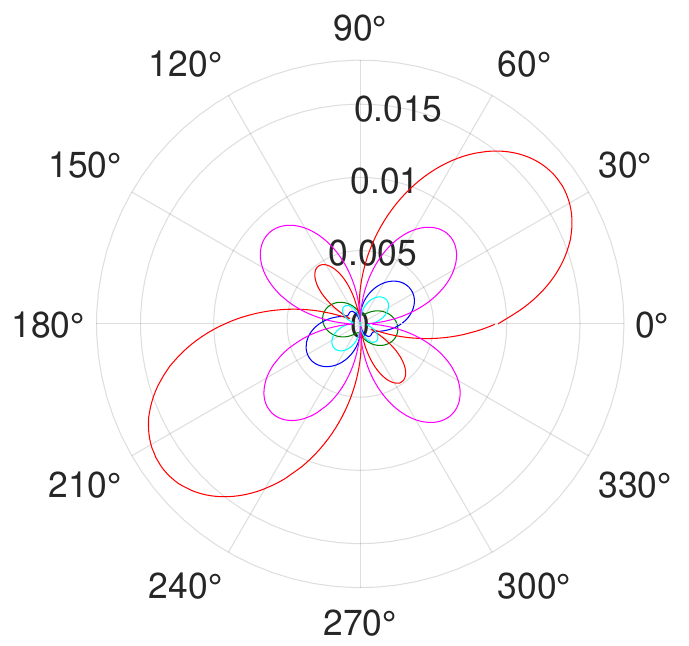}\vspace{.2cm}} & \parbox[c]{\linewidth}{\centering \vspace{.2cm}\includegraphics[scale=0.4]{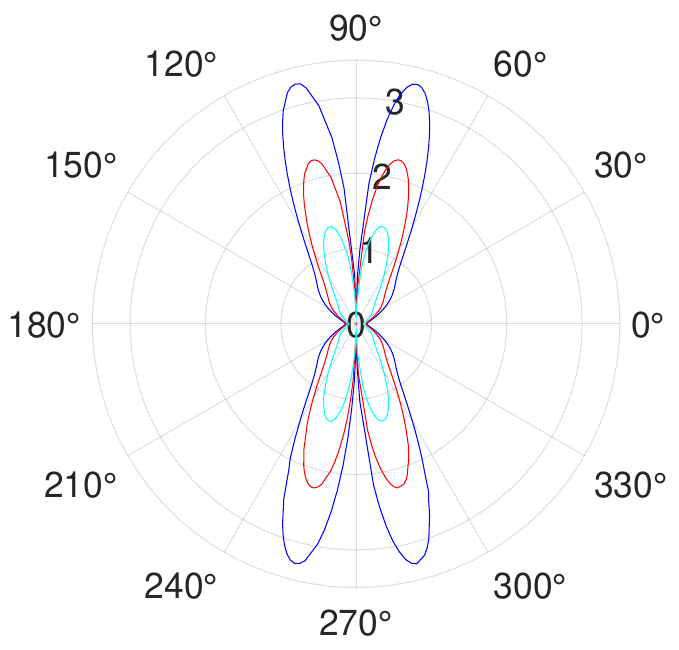}\vspace{.2cm}} & \parbox[c]{\linewidth}{\centering \vspace{0.2cm}\includegraphics[scale=0.4]{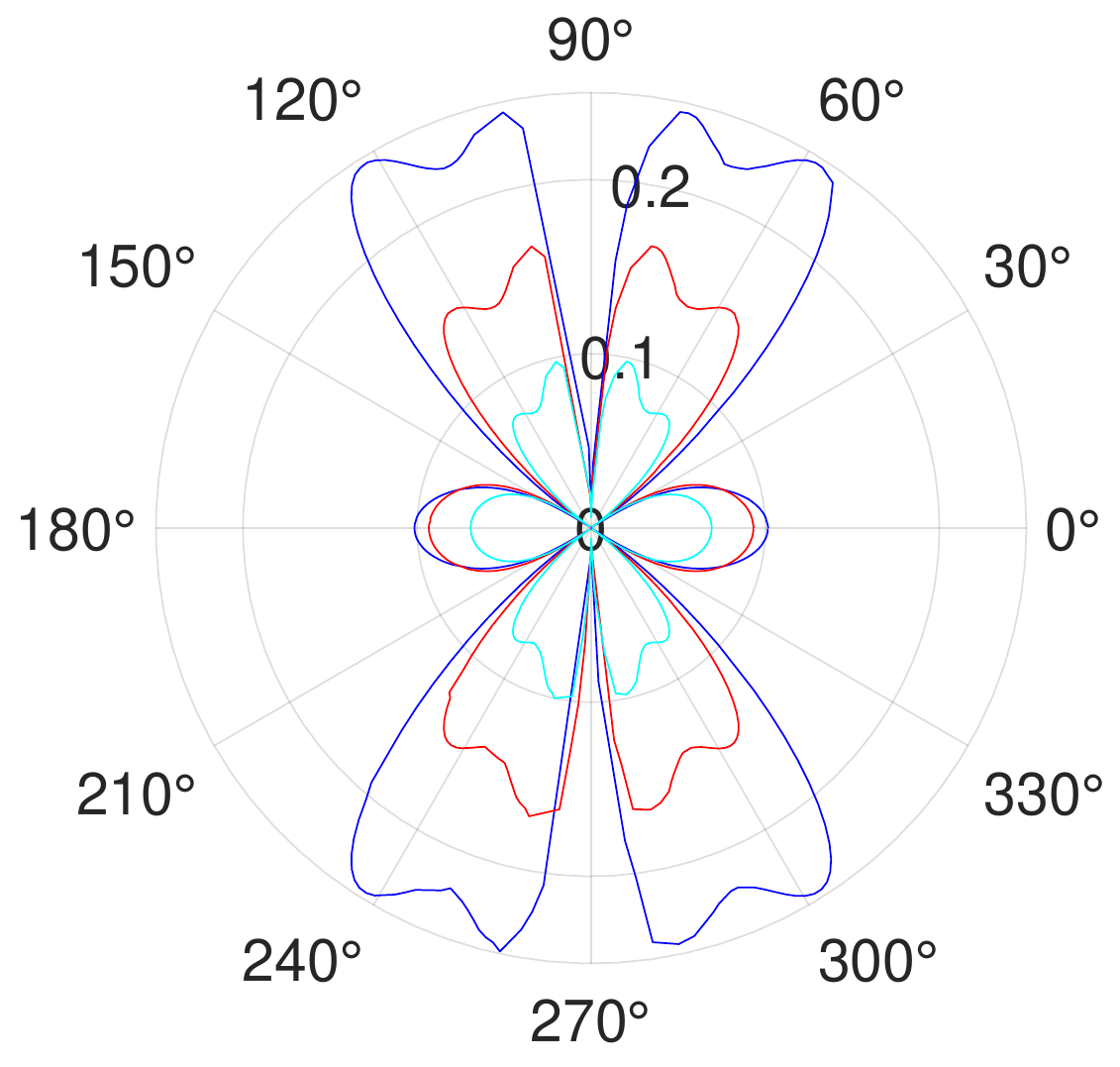}\vspace{.2cm}} \\
\hline
\end{tabular}
    \caption{Form factors for Mode 1 in Cavity 1 as a function of the incidence angle $\theta$. Rows make reference to the two different polarizations of the HFGW and columns make reference to the three different planes in which the incidence of the HFGW is studied. Each plates angle ($\alpha$) is illustrated with a different color: Green (\textcolor{verdeoscuro}{\rule[0.5ex]{0.5cm}{1pt}}) for $\alpha=0^{\circ}$, Blue (\textcolor{blue}{\rule[0.5ex]{0.5cm}{1pt}}) for $\alpha=30^{\circ}$, Red (\textcolor{red}{\rule[0.5ex]{0.5cm}{1pt}}) for $\alpha=45^{\circ}$, Cyan (\textcolor{cyan}{\rule[0.5ex]{0.5cm}{1pt}}) for $\alpha=60^{\circ}$ and Magenta (\textcolor{magenta}{\rule[0.5ex]{0.5cm}{1pt}}) for $\alpha=90^{\circ}$. In the $\tilde{\eta}_{+}$ \textit{XZ} plane and $\tilde{\eta}_{+}$ \textit{YZ} plane cases $\alpha=0^{\circ}$ and $\alpha=90^{\circ}$ cannot be seen since the coupling is much lower in comparison with the rest of the angles. The same occurs for $0^{\circ}$ in $\tilde{\eta}_{+}$ \textit{XY} plane case.}
    \label{Banda_UHF_modo_1}
\end{table*}
}

{
\begin{table*}[ht]
    %\captionsetup{type=figure}
    \centering
    \begin{tabular}{| c | >{\centering\arraybackslash}m{0.3\textwidth} | >{\centering\arraybackslash}m{0.3\textwidth} | >{\centering\arraybackslash}m{0.3\textwidth} |}
\cline { 2 - 4 } \multicolumn{1}{c|}{} & \vspace{0.2cm} \hspace{0.12cm} \Large \textit{XY} plane \vspace{0.1cm} & \vspace{0.2cm} \hspace{0.12cm} \Large \textit{XZ} plane \vspace{0.1cm} & \vspace{0.2cm} \hspace{0.12cm} \Large \textit{YZ} plane \vspace{0.1cm} \\
\hline \Large $\tilde{\eta}_{\times}$ & \parbox[c]{\linewidth}{\centering \vspace{.2cm}\includegraphics[scale=0.4]{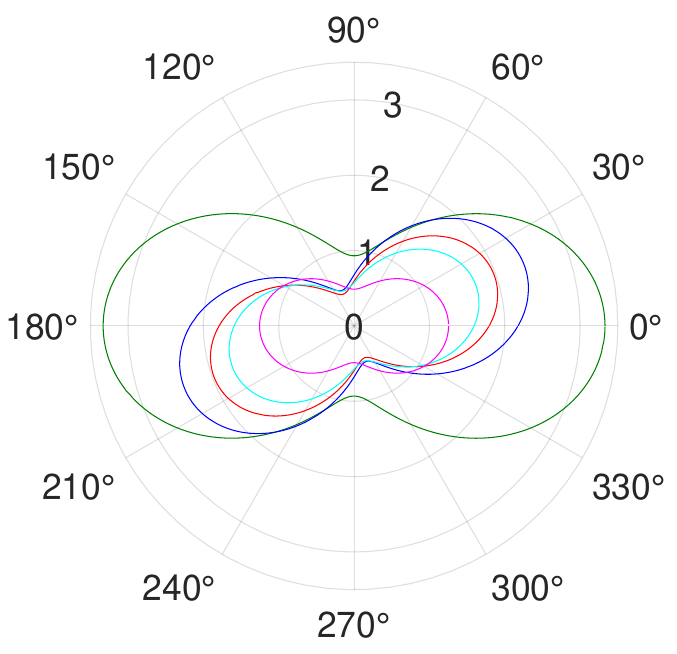}\vspace{.2cm}} & \parbox[c]{\linewidth}{\centering \vspace{.2cm}\includegraphics[scale=0.4]{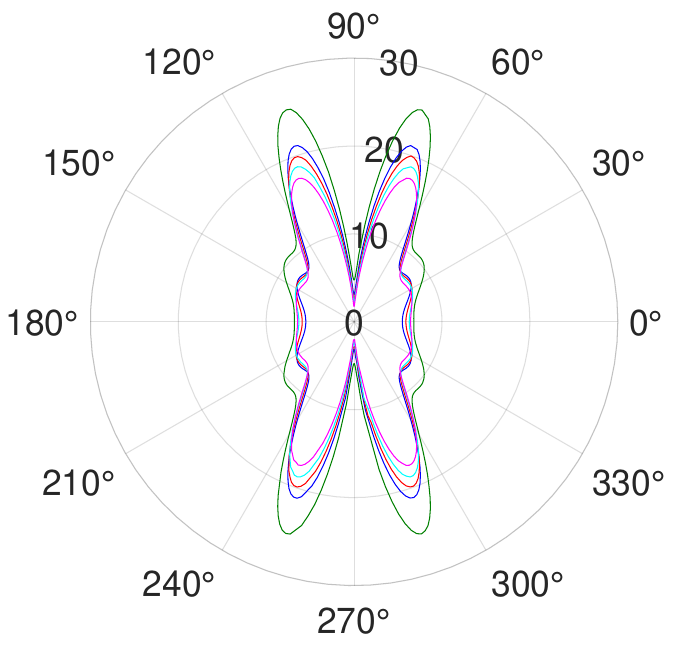}\vspace{.2cm}} & \parbox[c]{\linewidth}{\centering \vspace{.2cm}\includegraphics[scale=0.4]{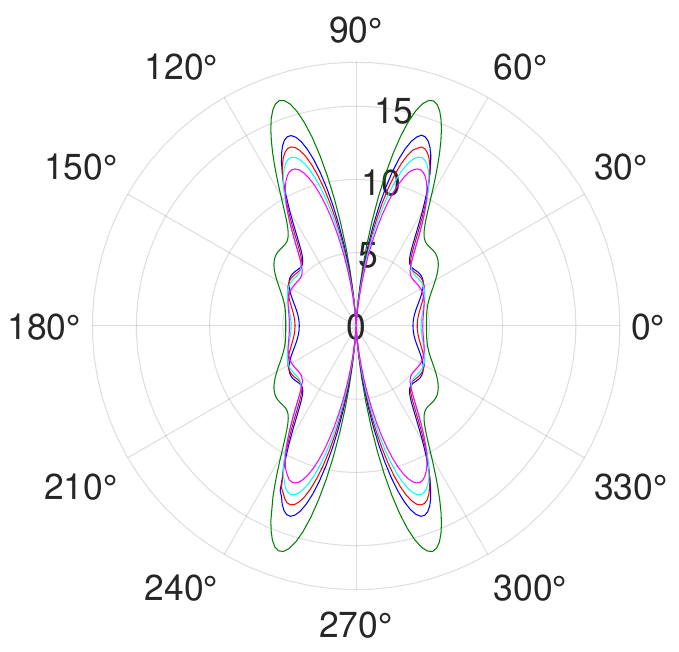}\vspace{.2cm}} \\
\hline \Large $\tilde{\eta}_{+}$ & \parbox[c]{\linewidth}{\centering \vspace{.2cm}\includegraphics[scale=0.4]{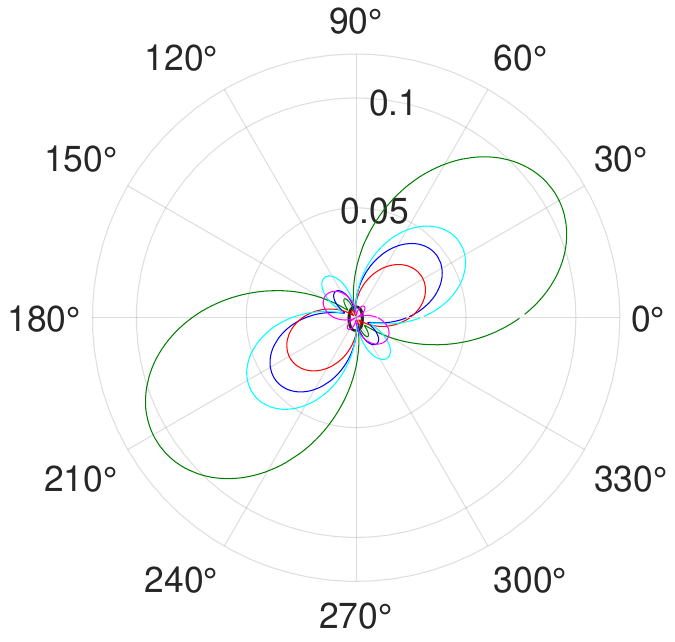}\vspace{.2cm}} & \parbox[c]{\linewidth}{\centering \vspace{.2cm}\includegraphics[scale=0.4]{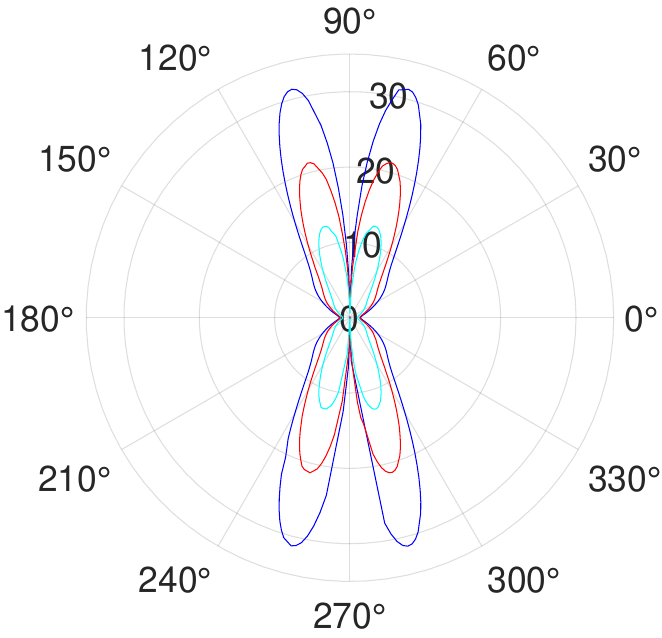}\vspace{.2cm}} & \parbox[c]{\linewidth}{\centering \vspace{0.2cm}\includegraphics[scale=0.4]{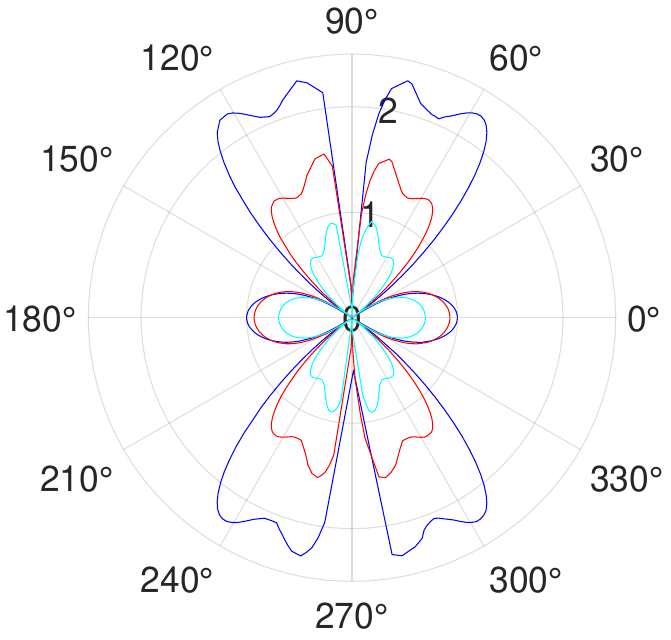}\vspace{.2cm}} \\
\hline
\end{tabular}
    \caption{Form factors for Mode 1 in Cavity 2 as a function of the incidence angle $\theta$. Rows make reference to the two different polarizations of the HFGW and columns make reference to the three different planes in which the incidence of the HFGW is studied. Each plates angle  ($\alpha$) is illustrated with a different color: Green (\textcolor{verdeoscuro}{\rule[0.5ex]{0.5cm}{1pt}}) for $\alpha=0^{\circ}$, Blue (\textcolor{blue}{\rule[0.5ex]{0.5cm}{1pt}}) for $\alpha=30^{\circ}$, Red (\textcolor{red}{\rule[0.5ex]{0.5cm}{1pt}}) for $\alpha=45^{\circ}$, Cyan (\textcolor{cyan}{\rule[0.5ex]{0.5cm}{1pt}}) for $\alpha=60^{\circ}$ and Magenta (\textcolor{magenta}{\rule[0.5ex]{0.5cm}{1pt}}) for $\alpha=90^{\circ}$. In the $\tilde{\eta}_{+}$ \textit{XZ} plane and $\tilde{\eta}_{+}$ \textit{YZ} plane cases $0^{\circ}$ and $\alpha=90^{\circ}$ cannot be seen since the coupling is much lower in comparison with the rest of the angles.}
    \label{Banda_S_Modo_1}
\end{table*}
}

{
\begin{table*}[ht]
    %\captionsetup{type=figure}
    \centering
    \begin{tabular}{| c | >{\centering\arraybackslash}m{0.3\textwidth} | >{\centering\arraybackslash}m{0.3\textwidth} | >{\centering\arraybackslash}m{0.3\textwidth} |}
\cline { 2 - 4 } \multicolumn{1}{c|}{} & \vspace{0.2cm} \hspace{0.12cm} \Large \textit{XY} plane \vspace{0.1cm} & \vspace{0.2cm} \hspace{0.12cm} \Large \textit{XZ} plane \vspace{0.1cm} & \vspace{0.2cm} \hspace{0.12cm} \Large \textit{YZ} plane \vspace{0.1cm} \\
\hline \Large $\tilde{\eta}_{\times}$ & \parbox[c]{\linewidth}{\centering \vspace{.2cm}\includegraphics[scale=0.4]{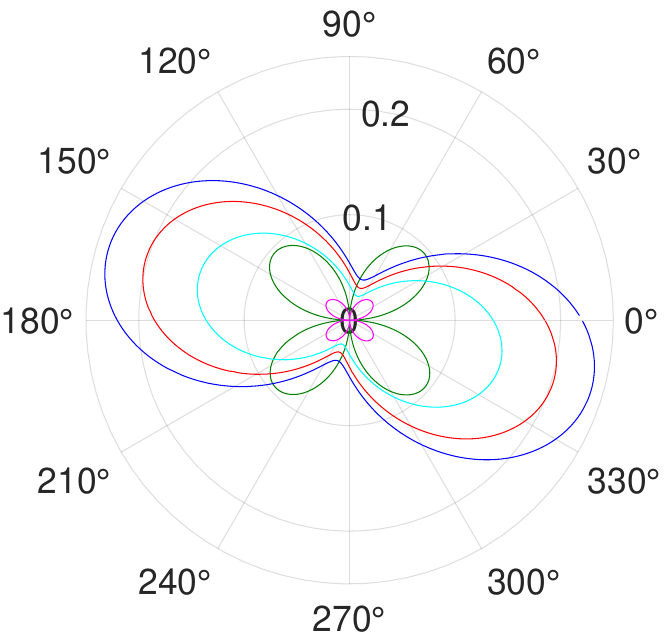}\vspace{.2cm}} & \parbox[c]{\linewidth}{\centering \vspace{.2cm}\includegraphics[scale=0.4]{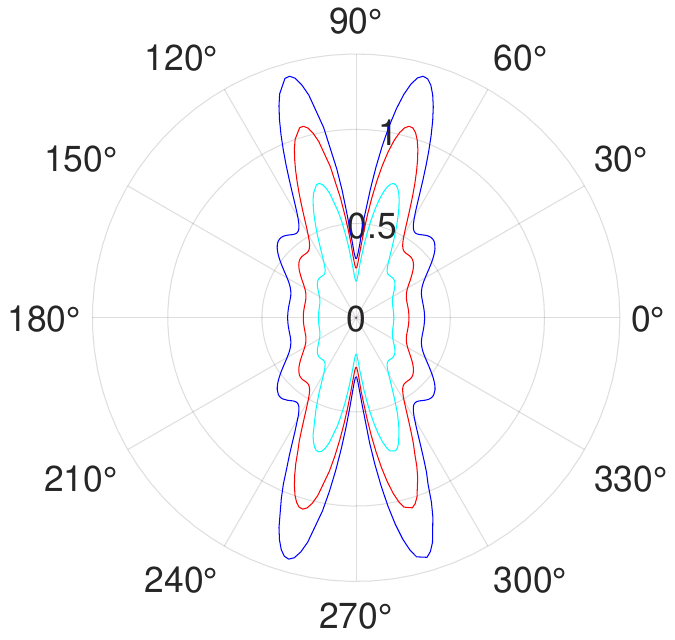}\vspace{.2cm}} & \parbox[c]{\linewidth}{\centering \vspace{.2cm}\includegraphics[scale=0.4]{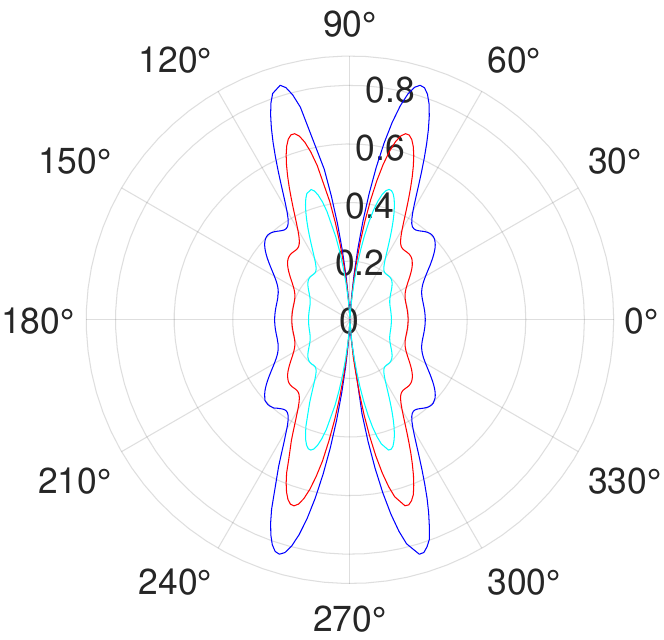}\vspace{.2cm}} \\
\hline \Large $\tilde{\eta}_{+}$ & \parbox[c]{\linewidth}{\centering \vspace{.2cm}\includegraphics[scale=0.4]{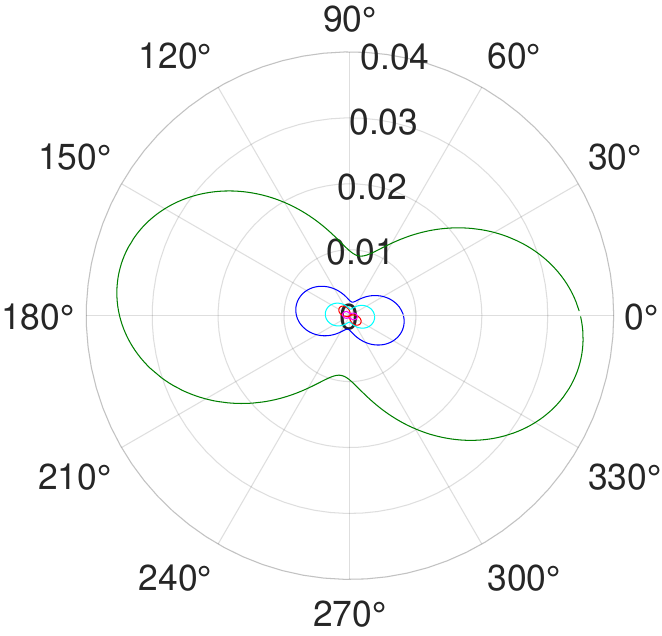}\vspace{.2cm}} & \parbox[c]{\linewidth}{\centering \vspace{.2cm}\includegraphics[scale=0.4]{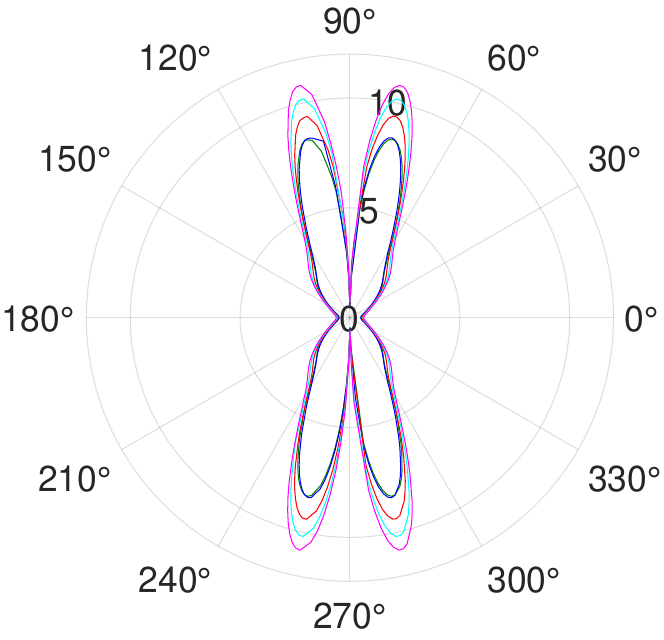}\vspace{.2cm}} & \parbox[c]{\linewidth}{\centering \vspace{0.2cm}\includegraphics[scale=0.4]{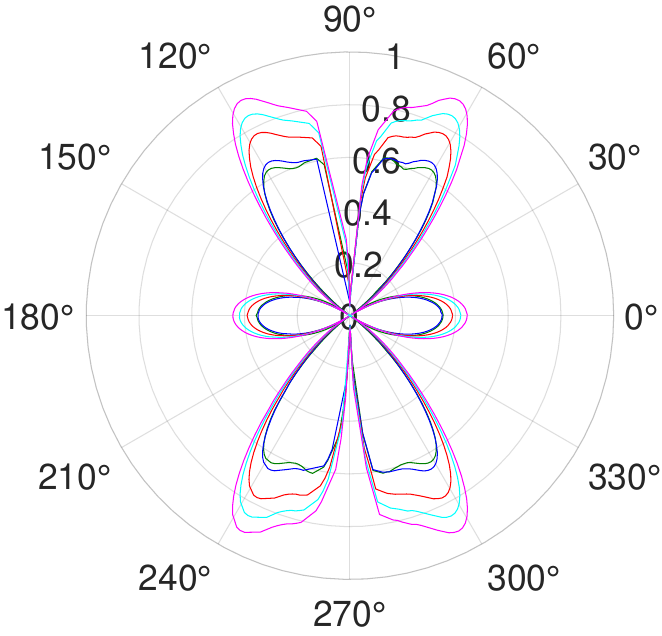}\vspace{.2cm}} \\
\hline
\end{tabular}
    \caption{Form factors for Mode 2 in Cavity 1 as a function of the incidence angle $\theta$. Rows make reference to the two different polarizations of the HFGW and columns make reference to the three different planes in which the incidence of the HFGW is studied. Each plates angle  ($\alpha$) is illustrated with a different color: Green (\textcolor{verdeoscuro}{\rule[0.5ex]{0.5cm}{1pt}}) for $\alpha=0^{\circ}$, Blue (\textcolor{blue}{\rule[0.5ex]{0.5cm}{1pt}}) for $\alpha=30^{\circ}$, Red (\textcolor{red}{\rule[0.5ex]{0.5cm}{1pt}}) for $\alpha=45^{\circ}$, Cyan (\textcolor{cyan}{\rule[0.5ex]{0.5cm}{1pt}}) for $\alpha=60^{\circ}$ and Magenta (\textcolor{magenta}{\rule[0.5ex]{0.5cm}{1pt}}) for $\alpha=90^{\circ}$. In the $\tilde{\eta}_{\times}$ \textit{XZ} plane and $\tilde{\eta}_{\times}$ \textit{YZ} plane cases $\alpha=0^{\circ}$ and $\alpha=90^{\circ}$ cannot be seen since the coupling is much lower in comparison with the rest of the angles. The same occurs with $\alpha=30^{\circ}$ and $\alpha=90^{\circ}$ form factor for $\tilde{\eta}_{+}$ \textit{XY} plane case.}
    \label{Banda_UHF_Modo_2}
\end{table*}
}

{
\begin{table*}[ht]
    %\captionsetup{type=figure}
    \centering
    \begin{tabular}{| c | >{\centering\arraybackslash}m{0.3\textwidth} | >{\centering\arraybackslash}m{0.3\textwidth} | >{\centering\arraybackslash}m{0.3\textwidth} |}
\cline { 2 - 4 } \multicolumn{1}{c|}{} & \vspace{0.2cm} \hspace{0.12cm} \Large \textit{XY} plane \vspace{0.1cm} & \vspace{0.2cm} \hspace{0.12cm} \Large \textit{XZ} plane \vspace{0.1cm} & \vspace{0.2cm} \hspace{0.12cm} \Large \textit{YZ} plane \vspace{0.1cm} \\
\hline \Large $\tilde{\eta}_{\times}$ & \parbox[c]{\linewidth}{\centering \vspace{.2cm}\includegraphics[scale=0.4]{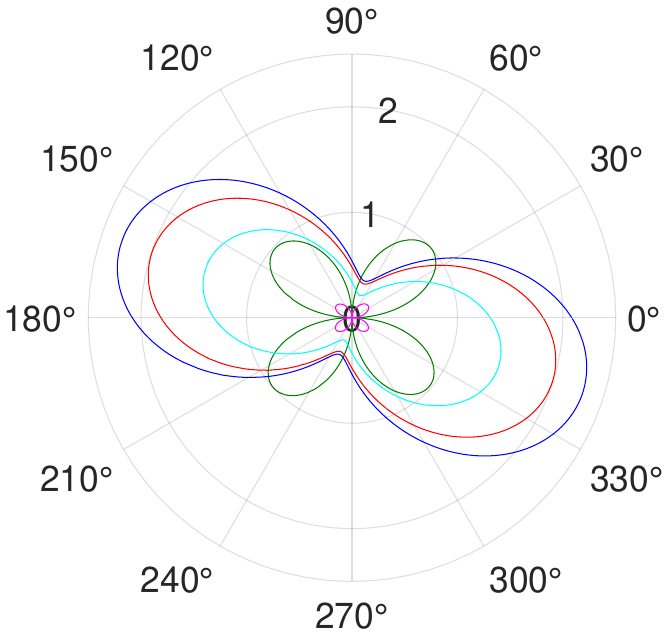}\vspace{.2cm}} & \parbox[c]{\linewidth}{\centering \vspace{.2cm}\includegraphics[scale=0.4]{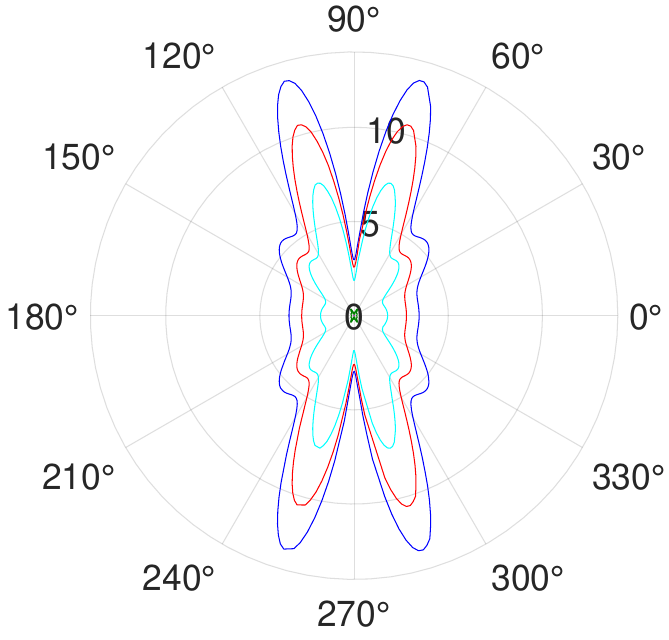}\vspace{.2cm}} & \parbox[c]{\linewidth}{\centering \vspace{.2cm}\includegraphics[scale=0.4]{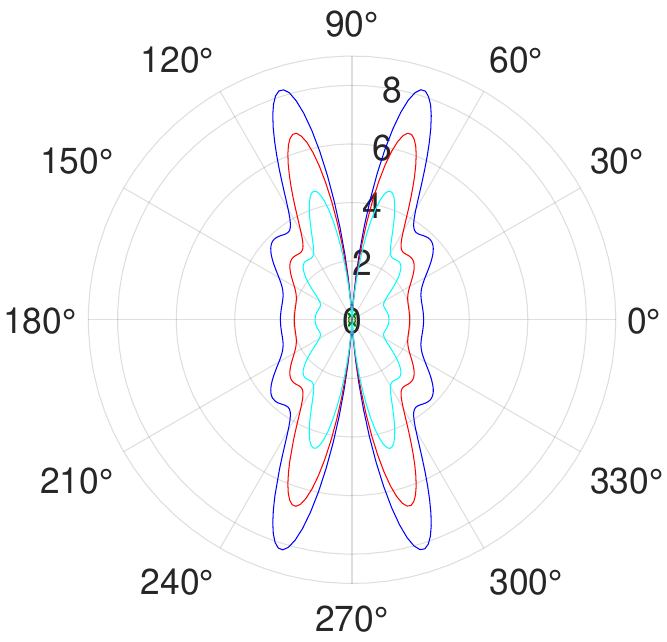}\vspace{.2cm}} \\
\hline \Large $\tilde{\eta}_{+}$ & \parbox[c]{\linewidth}{\centering \vspace{.2cm}\includegraphics[scale=0.4]{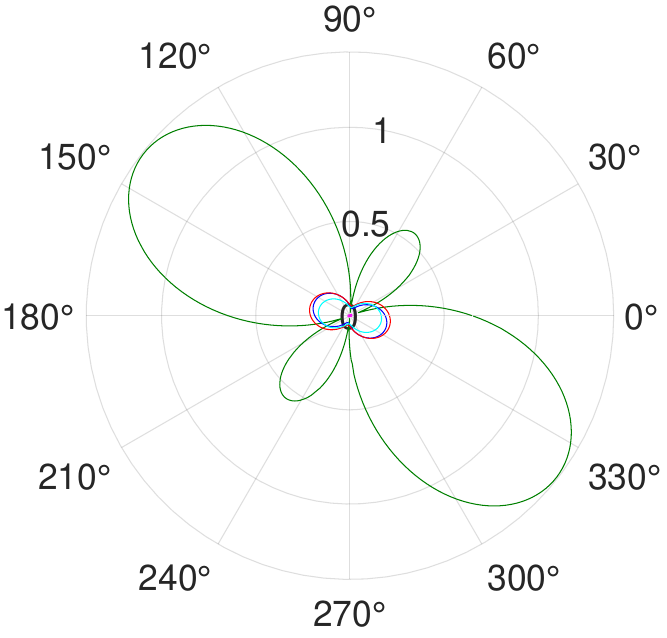}\vspace{.2cm}} & \parbox[c]{\linewidth}{\centering \vspace{.2cm}\includegraphics[scale=0.4]{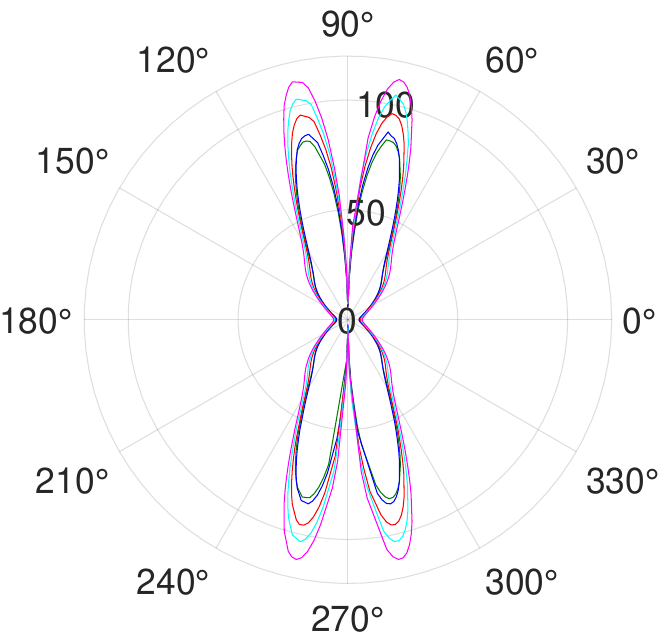}\vspace{.2cm}} & \parbox[c]{\linewidth}{\centering \vspace{0.2cm}\includegraphics[scale=0.4]{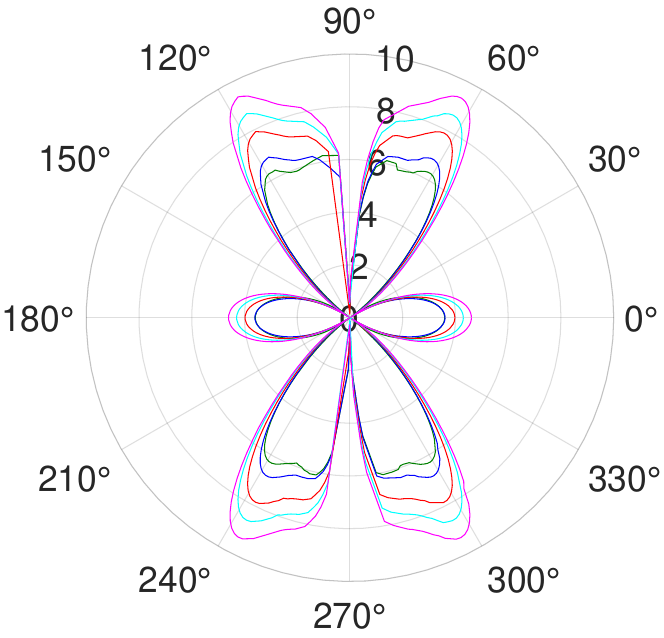}\vspace{.2cm}} \\
\hline
\end{tabular}
    \caption{Form factors for Mode 2 in Cavity 2 as a function of the incidence angle $\theta$. Rows make reference to the two different polarizations of the HFGW and columns make reference to the three different planes in which the incidence of the HFGW is studied. Each plates angle  ($\alpha$) is illustrated with a different color: Green (\textcolor{verdeoscuro}{\rule[0.5ex]{0.5cm}{1pt}}) for $\alpha=0^{\circ}$, Blue (\textcolor{blue}{\rule[0.5ex]{0.5cm}{1pt}}) for $\alpha=30^{\circ}$, Red (\textcolor{red}{\rule[0.5ex]{0.5cm}{1pt}}) for $\alpha=45^{\circ}$, Cyan (\textcolor{cyan}{\rule[0.5ex]{0.5cm}{1pt}}) for $\alpha=60^{\circ}$ and Magenta (\textcolor{magenta}{\rule[0.5ex]{0.5cm}{1pt}}) for $\alpha=90^{\circ}$. In the $\tilde{\eta}_{\times}$ \textit{XZ} plane and $\tilde{\eta}_{\times}$ \textit{YZ} plane cases $\alpha=0^{\circ}$ and $\alpha=90^{\circ}$ cannot be seen since the coupling is much lower in comparison with the rest of the angles.}
    \label{Banda_S_Modo_2}
\end{table*}
}
From this expression, when both the electric field and density current are perfectly aligned, the form factor will have maximum magnitude. However, the expression of $\vec{J}$ is relatively complex, making it non-trivial to analyse its effect \cite{PhysRevLett.129.041101}. In particular, even though the static magnetic field is oriented along the $y$-axis, the HFGW can couple to any cavity mode with different magnitudes of the coupling factor. Therefore, it is of interest to perform a numerical calculation of $\tilde{\eta}_{m_{+,\times}}$ making a sweep in  the angle of incidence of the HFGW (see Fig. \ref{fig:GW_incidence}), $\theta$, and fixing the remaining parameters to a given value. For this purpose, a simulation of the RF electromagnetic fields inside the cavity has been performed with the eigenmode solver of CST Studio Suite software \cite{CSTStudioSuite}.

The study is divided in the following cases: first, a given polarization of the HFGW ($+$ or $\times$) is selected; then, we study a given plane of incidence in order to define $\theta$ (defined with respect to the $z$-axis in the \textit{XZ}, \textit{YZ} incidence planes and to the $x$-axis in the \textit{XY} incidence plane); finally, the tuning plates angle ($\alpha$) is fixed to a given value, defined with respect to the $x$-axis. Results for the Mode 1 of Cavity 1 are shown in Table. \ref{Banda_UHF_modo_1}. Different values of $\alpha$ ($0^{\circ}$, $30^{\circ}$, $45^{\circ}$, $60^{\circ}$ and $90^{\circ}$) have been chosen in order to sweep the main region of interest in which the structure of the electric field is modified considerably. Any result of $\tilde{\eta}_{m_{+,\times}}$ associated to another different value of $\alpha$ can be interpolated from the results showed here. The external magnetostatic field is assumed uniform, constant and oriented towards $y$-axis in the calculations \footnote{Nevertheless, as can be seen in \cite{BABYIAXO_RADES}, since it has some distortions in the external part of the bore, the magnetic field of the BabyIAXO experiment is not completely uniform. This aspect is more relevant in Cavity 1 since its dimensions are higher}.
The result of the form factor varies considerably with $\theta$, generating the kind of diagrams observed in Tables~\ref{Banda_UHF_modo_1}-\ref{Banda_S_Modo_1}. The existence of specific values of $\theta$ for which the system is blind to the incident HFGW is a crucial aspect of this analysis. In order to avoid these blind spots, daily modulation of the HFGW signal can be taken into account since the incidence angle of the gravitational wave will change and so will the form factor pattern, always provided that the source is stationary. In addition, studying more cavity modes that differ in their electric field structure can be useful to avoid blind spots~\cite{HFGW_pablo}. An aspect associated with these angles of sensitivity drop is that their position in the polar plots does not vary considerably when changing the plates angle $\alpha$. Thus,  one may employ two or more cavities oriented towards different directions in order to have full sky coverage.

 A related aspect is the variation of the form factor $\tilde{\eta}$ with $\alpha$. There are cases where the system is more sensitive to an incident HFGW, as seen for instance in Tables~\ref{Banda_UHF_modo_1}-\ref{Banda_S_Modo_1} when varying $\alpha$ from $0^{\circ}$ (green) to $90^{\circ}$ (magenta). These changes can be as big as an order of magnitude. As a consequence, when studying the coupling between a HFGW and a resonant microwave cavity, the influence of the tuning system is of special relevance. Moreover, this aspect not only affects the RF modal electric field, but also $\vec{J}_{+,\times}$ due to its dependence with frequency. This can be seen when comparing results of Table \ref{Banda_S_Modo_1} with Table \ref{Banda_UHF_modo_1}, where it is clear that not only an order of magnitude has been gained, but also a change in the pattern of the $\tilde{\eta}_{+}$ \textit{XY} case can be seen.
%One aspect that has to be highlighted is the complex dependence of the gravitational wave current $\vec{J}$ with the frequency. This will generate that in some cases, even though the geometry of the cavity is the same, the result of the form factor will vary considerably since we are changing the resonant frequency of the cavity, i.e., the frequency of the HFGW that we are interested in studying. This can be seen together with Fig. \ref{Banda_S_Modo_1} when comparing results with the previous figure. Not only the order of magnitude of the results has changed in general, but also the $\tilde{\eta}_{+}$ \textit{XY} plane case has seen its pattern modified. As can be observed, an order of magnitude has been gained in each case when changing from Cavity 1 to Cavity 2, implying that the latter system is considerably more sensitive to the incidence of a HFGW than the first one.

The form factor for Mode 2 has also been computed. Results are shown in Tables \ref{Banda_UHF_Modo_2} and \ref{Banda_S_Modo_2}. As it is seen in Fig. \ref{fig:E_field_profile}, initially ($\alpha = 0^{\circ}$), this cavity mode has its RF electric field aligned with the $x$-axis, which implies a different effect on the form factor in the \textit{XY} incidence plane during the frequency tuning. The cases  $\alpha=0^{\circ}$ and $\alpha=90^{\circ}$ on $\tilde{\eta}_{\times}$ \textit{XZ} and \textit{YZ} planes are of special relevance due to the decrease of the form factor with respect to the rest of values of the angle $\alpha$, behaviour that was also seen in the earlier tables of Mode 1 but for the $+$ polarization.

The same aspects discussed for Mode 1 can be extrapolated for Mode 2: the presence of blind spots for given values of $\theta$ in certain polarizations and incidence planes; the drastic change in the results when varying the plates angle, seen for instance in the $\tilde{\eta}_{+}$ \textit{XY} plane case (Table \ref{Banda_UHF_Modo_2})%, where the results increase more than one order of magnitude of the $0^{\circ}$ (green) results
; and the change of the diagrams pattern not only between the two different cavities studied but also with respect to Mode~1.

The study of the form factor of Mode 2 allows us to  duplicate the scanning frequency range, due to the split evolution between modes 1 and 2 (see Fig. \ref{fig:freq_tuning}). Furthermore, it is key to assess the origin of the detected RF signal, as we now explain. Two situations are possible: both ports are excited; or only port 1 or port 2 is excited. The first case implies that the bandwidth of the detected RF signal is larger than the frequency gap between modes for a given angle of the plates. Since axions would not be able of exciting modes 1 and 2 at once due to its expected bandwidth \footnote{Excluding the possibility of two axions with a mass split corresponding to the one between both modes.}, the signal could have been generated by a HFGW. If only port 1 is excited, then a study of the daily modulation of the signal must be performed in order to distinguish the cosmological source. Finally, if port 2 is excited alone, a consideration has to be taken into account. When varying the plates angle the electromagnetic field is being distorted, and as a consequence, an axion would be able to excite the second Mode. Even though this coupling would be lower than the one of Mode 1, it could be even more significant than the coupling of a HFGW due to its expected low amplitude. With that, it can be concluded that for this case an study of the time modulation of the RF signal would be also required. 

Let us now be more explicit about the time variation of the signal, given its relevance to identify the kind of source behind the latter. For the axion case, the time modulation of the signal arises when considering the influence of the typical momentum  of the axion field $\vec k_{a}$, introducing the spatial frequency term $\vec{k}_{a}\cdot \vec{r}$, where $\vec{r}$ is the spatial coordinates vector of the cavity. Since $\vec{k}_{a}$ is proportional to the velocity with which we see axions from our laboratory reference frame, the consideration of this term translates into a tiny daily modulation of the signal, suppressed by velocity $v\sim10^{-3}c$ (being $c$ the speed of light in vacuum) of the axion galactic wind in our galaxy, as compared to the main RF signal  \cite{Knirck_2018} \cite{Igor_directional}. Higher masses of the axion (higher momentum) and cavities of size comparable to the de Broglie wavelength maximize the influence of this term. Regarding HFGWs, the profile of the signal's time modulation strongly depends on which type of source produces the wave. For instance, if the signal is produced by the collision between two primordial black holes, a chirping spectra is expected, and a signal duration which may be rather short \cite{Maggiore,Franciolini:2022htd}. Hence,  the total time that the signal couples in band with a resonant mode has to be considered, in particular with respect to the tuning system employed in the experiment. Recall that changing the angle $\alpha$ of the plates inside the cavity modifies its resonant frequency, allowing us to sweep over a frequency range. If the HFGW bandwidth is high enough, a merging source may generate a signal in the cavity for more than one angle of plates $\alpha$. The fundamental question is how much time we integrate for a given value of $\alpha$. One solution would be to allow the system for fast read-outs, where the frequency change when moving the plates from one angle to another is less than the frequency variation of the HFGW. However,  the mechanical tuning system can take a few minutes to move from one of the angles studied to another, and this possibility seems unlikely.  A more realistic strategy would be  working with an array of cavities tuned to different frequencies for a simultaneous scan over a frequency range. This case is also favoured, as the existence of chirping may be used as a way to confirm the origin of the signal, thus increasing detection prospects.
% But since the kind of HFGWs source mentioned is transient, this implies that when sweeping from one frequency to another we are loosing the capability of detecting a signal with the previous frequency that passes through our experiment
 %The latter case would be the ideal one since it would allow us to have arrays of cavities each one with different values of $\alpha$ between them. With that, during the duration of the HFGW event, each array of cavities detects different parts of the frequency range of the chirp signal. 
 
 Finally, superradiant clouds associated to new bosons of mass $m_b$ can produce a stationary signal in our detector. These are long-lasting phenomena around spinning black holes of mass $M_{BH}\sim (G m_b)^{-1}$ ($G$ is Newton's constant) whose duration is over thousands of years and generate HFGWs in the band $\sim m_b$ \cite{Brito:2015oca}.  Thus, the tuning system of our set up is ideal to search for this kind of sources. In this case, the rotation of the Earth would generate an $O(1)$ modification in the coefficients $\tilde \eta_{+,\times}$, from the relative rotation of $\vec k_{GW}$ with respect to the cavity. Another source of persistent GWs are the stochastic backgrounds generated by several sources, possibly of primordial origin~\cite{Aggarwal:2020olq}. The treatment of the signals induced by GW backgrounds goes beyond the current analysis. 
 %To distinguish between these two possibilities, as we discussed, one can use different modes with very close frequencies, or, simpler, the fact that the rotation of the Earth modulates the signal for HFGWs.

%%%%%%%%%%%%%%%%%%%%%%%%%%%%%%%%%%%%%%%%%%%%%%%%%%%%%%%%%%%%
\subsection{Sensitivity estimation}
%%%%%%%%%%%%%%%%%%%%%%%%%%%%%%%%%%%%%%%%%%%%%%%%%%%%%%%%%%%%

At this point, a sensitivity estimation of the experiment can be made. 
%In this work, multiple situations have been examined; only the predicted sensitivity of the experiment is computed for specific fixed values of the parameters.
The expected power generated by the HFGW inside the cavity can be expressed in SI units as \cite{Berlin:2021txa}:
\begin{equation}
    P_{\mathrm{sig}} = \frac{\varepsilon_{0}}{2}\,Q_{\mathrm{eff}}\,\omega_{c}^{3}\,V^{5/3}\left(\tilde{\eta}\,h_{0}\,B_{0}\right)^2,
\end{equation}
where $\varepsilon_{0}$ is the vacuum electric permittivity; $Q_{\mathrm{eff}}$ is an effective quality factor; $\omega_{c}$ is the angular frequency of the HFGW; $V$ is the cavity volume; $h_{0}$ is the gravitational wave amplitude; and $B_{0}$ is the external static magnetic field. Since we do not have a specified source of interest, the value adopted by $Q_{\mathrm{eff}}$ will be the loaded cavity quality factor $Q_L$. Of course, when making this decision it is assumed that $Q_{\mathrm{GW}}$ $\gg$ $Q_{L}$ (where $Q_{\mathrm{GW}}$ is the quality factor of the GW), being in a situation in which the signal has been detected only in one port.
From here, we can obtain the value of $h_{0}$ that can be reached taking into account Dicke's radiometer equation for the noise power \cite{Pozar}, where $T_{\mathrm{sys}}$ is the sum of the noise temperature of the cavity, i.e. its physical temperature, and the noise temperature added by the readout chain, which is approximately that of the first amplifier; $k_{B}$ is the Boltzmann constant; $t_{\mathrm{int}}$ is the integration time; and $\Delta \nu$ is the frequency detection bandwidth:
\begin{equation}
    \mathrm{SNR} = \frac{P_{\mathrm{d}}}{k_{B}\,T_{\mathrm{sys}}}\sqrt{\frac{t_{\mathrm{int}}}{\Delta\nu}},
    \label{SNR}
\end{equation}
where $P_{\mathrm{d}}$ $=$ $\frac{\beta}{1+\beta}P_{\mathrm{sig}}$ is the detected power (recall that $\beta$ is the coupling coefficient \cite{Pozar}). By fixing $\mathrm{SNR}$ $>$ $3$, %(i.e., we assume that our experiment can only be sensitive to a signal that accomplishes this $\mathrm{SNR}$):
\\
\begin{equation}
    h_{0} > \frac{1}{\tilde{\eta}\,B_0}\left(\frac{\Delta \nu}{t_{\mathrm{\mathrm{int}}}}\right)^{1/4}\left(\frac{1+\beta}{\beta}\frac{6 \ k_{B}\,T_{\mathrm{sys}}}{\varepsilon_{0}\,Q_{\mathrm{eff}}\,\omega_{c}^3\,V^{5/3}}\right)^{1/2}.
\end{equation}
Employing the values for the case of Cavity 1 with Mode 1, $\alpha$ $=$ $0^{\circ}$ in \textit{XY} plane for $\times$ polarization and $\theta$ $=$ $0^{\circ}$ (which corresponds to the maximum form factor for this case), we obtain the sensitivity expectation,
\begin{widetext}
\begin{equation}
\begin{aligned}
    h_{0} > 9.38 \ \cdot \ 10^{-22}&\left(\frac{\frac{1+\beta}{\beta}}{3/2}\right)^{1/2}\left(\frac{0.3356}{\tilde{\eta}}\right)\left(\frac{2\ \mathrm{T}}{B_{0}}\right)\left(\frac{\Delta\nu}{1.03 \ \mathrm{kHz}}\frac{2\ \mathrm{min}}{t_{\mathrm{int}}}\right)^{1/4}\left(\frac{T_{\mathrm{sys}}}{4.6 \ \mathrm{K}}\frac{96010}{Q_{\mathrm{eff}}}\right)^{1/2} \\
    & \times \left(\frac{1.9 \cdot 10^{9} \ \mathrm{rad/s}}{\omega_{c}}\right)^{3/2}\left(\frac{1.1898\ \mathrm{m^{3}}}{V}\right)^{5/6}. 
    \label{eq:sensitivity_UHF_band}
\end{aligned}
\end{equation}
\end{widetext}
The value employed for the coupling factor, $\beta$ $=$ $2$, is commonly used in axion haloscope experiments \cite{detection_rate_JCAP}. Since the form factors obtained in this work vary up to two orders of magnitude with respect to the one chosen for this calculation (as we have taken an average value, this variation can act either enhance or suppress the RF signal by an order of magnitude), this must be considered just an illustrative example of the achievable sensitivity that the RADES-BabyIAXO setup can reach. 

The sensitivity of \eqref{eq:sensitivity_UHF_band} may be improved in a number of ways: the integration time of 2 minutes may be extended, in particular considering the possibility of stationary sources of HFGW; lower temperatures inside the cryostat will be achievable with \textit{DarkQuantum}, reaching temperatures up to tens of mK by the development of DL technologies and quantum limited amplification with SQUIDs; the considered unloaded quality factor of the cavity is the cryogenic one for pure Copper, but this value can be improved by coating the cavity with superconducting materials such as Rare-Earth Barium Copper Oxide (REBCO) or $\mathrm{Nb_{3}Sn}$ \cite{superconducting_cavities}.

The sensitivity expectations for both cavities and modes are shown in Table \ref{tab:sensitivity_expectation}. In each case, the employed value for the form factor corresponds to the maximum one of the $\tilde{\eta}_{\times}$ \textit{XY} plane cases.
The achievable sensitivities are still far away from the  expected sources~\cite{FLASH}. However, it should be remarked that these are the first steps in a dynamical field of research. Conservative values of the experimental parameters have been employed for our estimates, and  there is much room for improvement of the experimental parameters in the incoming years.\\

\begin{table}[h]
\centering
\begin{tabular}{| c | c | c |}
\hline
 & Mode 1 & Mode 2\\
\hline
Cavity 1 & 0.94 & 4.00 \\ 
\hline
Cavity 2 & 5.45 & 23.01 \\

\hline
\end{tabular}
\caption{Sensitivity expectation of both cavities and modes for $h_{0}/ 10^{-21}$.}
\label{tab:sensitivity_expectation}
\end{table}

\subsection{Scanning rate}
\label{sec:scanning_rate}
We can now perform a study of the scanning rate as is commonly made in axion experiments but particularizing it for the HFGW case. To deduce the detection rate expression, a similar approach to the one in  \cite{universe_RADES,detection_rate_JCAP} has been followed, obtaining :
\begin{widetext}
\begin{equation}
\begin{aligned}
    \frac{\mathrm{d}\nu}{\mathrm{d}t} = \left(\frac{\frac{\varepsilon_{0}}{2}\,\omega_{c}^3\,V^{5/3}\,\tilde{\eta}^2\,h_{0}^2\,B_{0}^2}{\mathrm{SNR}\ k_{B}\,T_{\mathrm{cav}}}\right)^2\left(\frac{\frac{\beta}{\left(1+\beta\right)}}{\frac{4\beta}{\left(1 + \beta\right)^2} + \lambda}\right)^2\frac{Q_{\mathrm{GW}}^{3}\,Q_{L}}{\left(Q_{\mathrm{GW}} + Q_{L}\right)^2},
    \label{eq:scanning_rate}
\end{aligned}
\end{equation}
\end{widetext}
where $\lambda$ $=$ $T_{\mathrm{amp}}/T_{\mathrm{cav}}$, being $T_{\mathrm{amp}}$ the temperature of the amplification stage and $T_{\mathrm{cav}}$ the physical temperature of the cavity.Using the  same parameters employed in the sensitivity calculation section, this time in Eq. (\ref{eq:scanning_rate}), we find the results shown in Table \ref{tab:scanning_rate}.
Even though these numbers have been derived for the mentioned experimental parameters, we must remark that they can vary considerably since, for instance, $\tilde{\eta}$ depends on the frequency of the incident signal. % As a result, these values must be taken as indications of the possible scanning rates. 

\begin{table}[ht]
\centering
\begin{tabular}{| c | c | c |}
\hline
 & Mode 1 & Mode 2 \\
\hline
Cavity 1 & 1.88 & 3.68 \\ 
\hline
Cavity 2 & 464.16 & 937.83 \\

\hline
\end{tabular}
\caption{Scanning rate of both cavities and modes in Hz/s.}
\label{tab:scanning_rate}
\end{table}

\begin{figure}[h]
    \centering
    \includegraphics[width=1\linewidth]{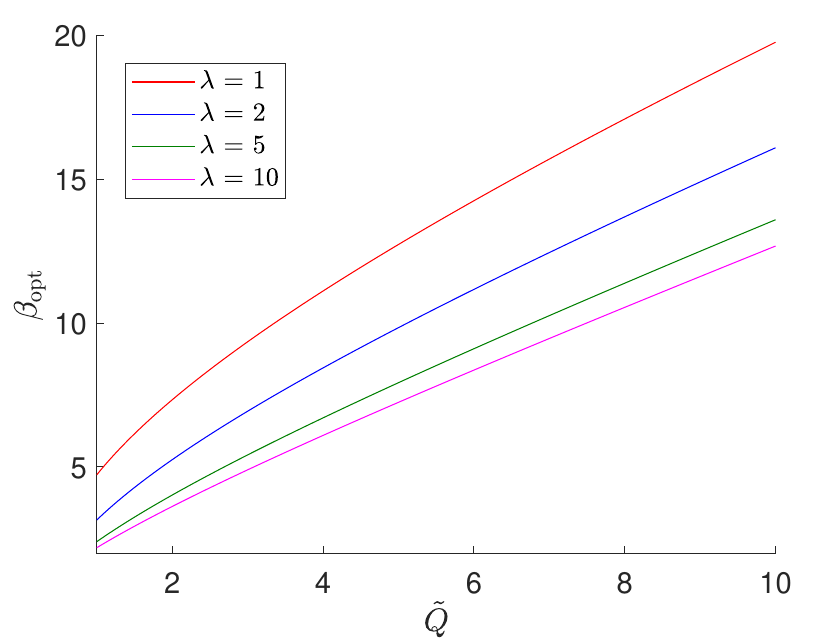}
    \caption{Behaviour of the optimum coupling factor $\beta_{\mathrm{opt}}$ when varying $\lambda$ and $\tilde{Q}$.}
    \label{fig:beta_vs_Q}
\end{figure}

The optimal $\beta$ in order to maximize this figure of merit can now be deduced, taking into account that $Q_{L}$ $=$ $\frac{Q_{0}}{1+\beta}$, where $Q_{0}$ is the unloaded quality factor of the cavity. The extreme of (\ref{eq:scanning_rate}) will be achieved for the solution of 
\begin{equation}
\begin{aligned}
    \lambda\beta^{4} & + \left(-4 + 2\lambda - \tilde{Q}\lambda\right)\beta^{3} \\
    & + \left(8 - 12\tilde{Q} + \lambda - 4\tilde{Q}\lambda\right)\beta^{2} - 5\tilde{Q}\lambda\beta - 2\tilde{Q}\lambda = 0,\nonumber
    \label{eq:4th_degree}
\end{aligned}
\end{equation}
where $\tilde{Q} = \frac{Q_{0}}{Q_{\mathrm{GW}}} + 1$. The result strongly depends on the values that $\lambda$ and $\tilde{Q}$ adopt as it can bee seen in Fig. \ref{fig:beta_vs_Q}. In  dark matter axion detection with haloscopes, one sets $\beta_{\mathrm{opt}}$ $=$ $2$, which is obtained when considering $\lambda$ $\gg$ $1$ and $\tilde{Q}$ $\approx$ $1$. However, for the case of BabyIAXO, due to quantum limited amplification with SQUIDs and other technologies, in the incoming years it would be able to decrease $\lambda$ and increase $\tilde{Q}$. In this way, representative values for this parameters taking into account these expectations in quantum technology would be $\lambda$ $=$ $5$ and $\tilde{Q}$ $=$ $2$, thus obtaining an optimum coupling $\beta_{\mathrm{opt}}$ $=$ $4$.

%%%%%%%%%%%%%%%%%%%%%%%%%%%%%%%%%%%%%%%%%%%%%%%%%%%%%%%%%%%%%%%%%%%%%%%%%%%%%%%
\section{Conclusions}
%%%%%%%%%%%%%%%%%%%%%%%%%%%%%%%%%%%%%%%%%%%%%%%%%%%%%%%%%%%%%%%%%%%%%%%%%%%%%%%

This article examines the possibilities of using RADES-BabyIAXO haloscope and a scaled version for higher frequencies as HFGWs detectors. For that, a calculation of the form factor between HFGW and two RADES-BabyIAXO cavities has been performed, analogously to the dark matter axion case outlined in \cite{HFGW_pablo}. The complexity of the electric current density generated by the HFGW  implies the coupling to a different set of modes supported by the cavity. Two orthonormal quasi-TE$_{111}$ modes have been studied, obtaining different results for the two HFGW polarizations, as well as for different planes and angles of incidence. A novel approach to the HFGWs detection has been proposed in this work, since it is the first time that the simultaneous detection of axion and HFGWs has been studied. In addition, a realistic set-up has been considered by incorporating the presence of tuning plates and coupling ports inside the cavity.

We have shown the relevance of this tuning system by observing the drastic change in the form factor when varying the plates angle. In addition, a sensitivity estimation of both cavities has been calculated, obtaining for both modes values of $h_{0}$ $\sim$ $10^{-21}$ for Cavity 1 and of $h_{0}$ $\sim$ $10^{-20}$ for Cavity 2 (cf. Table~\ref{tab:sensitivity_expectation}). These values could be considerably improved via optimization of the experimental parameters, with the final goal of achieving sensitivities closer to that of possible RF signals. Moreover, a qualitative study on how to distinguish a signal generated by a HFGW or an axion has been developed, concluding that RADES-BabyIAXO cavities are capable of distinguish both cases, in the case of stationary signals. The study of the form factor has been made for two orthonormal modes of both cavities, concluding that these two modes can be used for HFGWs detection and, therefore, achieve a duplication of the scanning frequency range. Our study can be easily expanded to other higher-order resonant modes, provided that the cavity contains the appropriate probes (loops or monopoles) for these modes. 
Moreover,  the scanning rate has been derived in Eq.~\eqref{eq:scanning_rate}, following a similar approach to that of the axion case. This allows to compute the optimal coupling factor $\beta$ maximizing the detection rate. The ideal HFGW scenario for the RADES-BabyIAXO setup are stationary sources, as superradiant black holes or stochastic GWs backgrounds (the latter is not considered in this study). Nevertheless, transient waves can also be detected even though this depends on the bandwidth of the HFGW.

In conclusion, BabyIAXO has potential to be employed as a HFGWs detector, while still being used as a dark matter axions detector. Regarding future directions, the employment of these cavities for an up-down conversion-type detector can be considered, similarly to \cite{Berlin:2023grv,australianos}. In these set-ups,  the resonant cavity is feeded by a pump mode and the presence of an axion or a HFGW can generate a signal in a readout mode. As the frequency gap  between Mode 1 and Mode 2 can be small \footnote{This is seen in  Fig.~\ref{fig:freq_tuning}. However, via a optimization of the cavity dimensions this frequency gap can be set to any desired value.},  the up-down conversion may be useful to explore HFGWs of frequencies below those considered in this work. By changing the plates angle, scanning to higher frequencies can be done since the gap between modes increases. Thus, RADES-BabyIAXO setup has a high capacity to be also considered as an up-down conversion detector.

\section{\label{acknowledgments} Acknowledgments}
This work was performed within the RADES group. We thank our colleagues for their support, specially Camilo García Cely for providing us with the currents expressions and his useful comments. Thanks also to Pablo Martín Luna for his valuable comments and discussions. The research leading to these results has received funding from the Spanish Ministry of Science and Innovation with the projects PID2020-115845GB-I00/AEI/10.13039/501100011033, PID2022-137268NBC53 and PID2022-137268NA-C55, funded by MICIU/AEI/10.13039/501100011033/ and by “ERDF/EU”. IFAE is partially funded by the CERCA program of the Generalitat de Catalunya. J. Reina-Valero has the support of "Plan de Recuperación, Transformación y Resiliencia (PRTR) 2022 (ASFAE/2022/013)", founded by Conselleria d'Innovació, Universitats, Ciència i Societat Digital from Generalitat Valenciana, and NextGenerationEU from European Union. D. Blas acknowledges the support from the Departament de Recerca i Universitats from Generalitat de Catalunya to the Grup de Recerca 00649 (Codi: 2021 SGR 00649). IGI acknowledges support from the  European Union’s Horizon 2020 research and innovation programme under the European Research Council (ERC) grant agreement ERC-2017-AdG788781 (IAXO+), as well as from the “European Union NextGenerationEU/PRTR” (Planes complementarios, Programa de Astrof\'isica y F\'isica de Altas Energías). 

\bibliography{references}

\providecommand{\noopsort}[1]{}\providecommand{\singleletter}[1]{#1}%
\begin{thebibliography}{29}%
\makeatletter
\providecommand \@ifxundefined [1]{%
 \@ifx{#1\undefined}
}%
\providecommand \@ifnum [1]{%
 \ifnum #1\expandafter \@firstoftwo
 \else \expandafter \@secondoftwo
 \fi
}%
\providecommand \@ifx [1]{%
 \ifx #1\expandafter \@firstoftwo
 \else \expandafter \@secondoftwo
 \fi
}%
\providecommand \natexlab [1]{#1}%
\providecommand \enquote  [1]{``#1''}%
\providecommand \bibnamefont  [1]{#1}%
\providecommand \bibfnamefont [1]{#1}%
\providecommand \citenamefont [1]{#1}%
\providecommand \href@noop [0]{\@secondoftwo}%
\providecommand \href [0]{\begingroup \@sanitize@url \@href}%
\providecommand \@href[1]{\@@startlink{#1}\@@href}%
\providecommand \@@href[1]{\endgroup#1\@@endlink}%
\providecommand \@sanitize@url [0]{\catcode `\\12\catcode `\$12\catcode `\&12\catcode `\#12\catcode `\^12\catcode `\_12\catcode `\%12\relax}%
\providecommand \@@startlink[1]{}%
\providecommand \@@endlink[0]{}%
\providecommand \url  [0]{\begingroup\@sanitize@url \@url }%
\providecommand \@url [1]{\endgroup\@href {#1}{\urlprefix }}%
\providecommand \urlprefix  [0]{URL }%
\providecommand \Eprint [0]{\href }%
\providecommand \doibase [0]{https://doi.org/}%
\providecommand \selectlanguage [0]{\@gobble}%
\providecommand \bibinfo  [0]{\@secondoftwo}%
\providecommand \bibfield  [0]{\@secondoftwo}%
\providecommand \translation [1]{[#1]}%
\providecommand \BibitemOpen [0]{}%
\providecommand \bibitemStop [0]{}%
\providecommand \bibitemNoStop [0]{.\EOS\space}%
\providecommand \EOS [0]{\spacefactor3000\relax}%
\providecommand \BibitemShut  [1]{\csname bibitem#1\endcsname}%
\let\auto@bib@innerbib\@empty
%</preamble>
\bibitem [{\citenamefont {Abbott}\ \emph {et~al.}(2016)\citenamefont {Abbott} \emph {et~al.}}]{LIGOScientific:2016aoc}%
  \BibitemOpen
  \bibfield  {author} {\bibinfo {author} {\bibfnamefont {B.~P.}\ \bibnamefont {Abbott}} \emph {et~al.} (\bibinfo {collaboration} {LIGO Scientific, Virgo}),\ }\bibfield  {title} {\bibinfo {title} {{Observation of Gravitational Waves from a Binary Black Hole Merger}},\ }\href {https://doi.org/10.1103/PhysRevLett.116.061102} {\bibfield  {journal} {\bibinfo  {journal} {Phys. Rev. Lett.}\ }\textbf {\bibinfo {volume} {116}},\ \bibinfo {pages} {061102} (\bibinfo {year} {2016})},\ \Eprint {https://arxiv.org/abs/1602.03837} {arXiv:1602.03837 [gr-qc]} \BibitemShut {NoStop}%
\bibitem [{\citenamefont {Aggarwal}\ \emph {et~al.}(2021)\citenamefont {Aggarwal} \emph {et~al.}}]{Aggarwal:2020olq}%
  \BibitemOpen
  \bibfield  {author} {\bibinfo {author} {\bibfnamefont {N.}~\bibnamefont {Aggarwal}} \emph {et~al.},\ }\bibfield  {title} {\bibinfo {title} {{Challenges and opportunities of gravitational-wave searches at MHz to GHz frequencies}},\ }\href {https://doi.org/10.1007/s41114-021-00032-5} {\bibfield  {journal} {\bibinfo  {journal} {Living Rev. Rel.}\ }\textbf {\bibinfo {volume} {24}},\ \bibinfo {pages} {4} (\bibinfo {year} {2021})},\ \Eprint {https://arxiv.org/abs/2011.12414} {arXiv:2011.12414 [gr-qc]} \BibitemShut {NoStop}%
\bibitem [{\citenamefont {Gertsenshtein}(1962)}]{gertsenshtein1962wave}%
  \BibitemOpen
  \bibfield  {author} {\bibinfo {author} {\bibfnamefont {M.}~\bibnamefont {Gertsenshtein}},\ }\bibfield  {title} {\bibinfo {title} {Wave resonance of light and gravitional waves},\ }\href@noop {} {\bibfield  {journal} {\bibinfo  {journal} {Sov Phys JETP}\ }\textbf {\bibinfo {volume} {14}},\ \bibinfo {pages} {84} (\bibinfo {year} {1962})}\BibitemShut {NoStop}%
\bibitem [{\citenamefont {Gertsenshtein}\ and\ \citenamefont {Pustov\u{o}it}(1962)}]{gertsenshtein_1962}%
  \BibitemOpen
  \bibfield  {author} {\bibinfo {author} {\bibfnamefont {M.~E.}\ \bibnamefont {Gertsenshtein}}\ and\ \bibinfo {author} {\bibfnamefont {V.~I.}\ \bibnamefont {Pustov\u{o}it}},\ }\bibfield  {title} {\bibinfo {title} {On the detection of low frequency gravitational waves},\ }\href@noop {} {\bibfield  {journal} {\bibinfo  {journal} {J. Exptl. Theoret. Phys.}\ }\textbf {\bibinfo {volume} {43}} (\bibinfo {year} {1962})}\BibitemShut {NoStop}%
\bibitem [{\citenamefont {Berlin}\ \emph {et~al.}(2022)\citenamefont {Berlin}, \citenamefont {Blas}, \citenamefont {Tito~D'Agnolo}, \citenamefont {Ellis}, \citenamefont {Harnik}, \citenamefont {Kahn},\ and\ \citenamefont {Sch\"utte-Engel}}]{Berlin:2021txa}%
  \BibitemOpen
  \bibfield  {author} {\bibinfo {author} {\bibfnamefont {A.}~\bibnamefont {Berlin}}, \bibinfo {author} {\bibfnamefont {D.}~\bibnamefont {Blas}}, \bibinfo {author} {\bibfnamefont {R.}~\bibnamefont {Tito~D'Agnolo}}, \bibinfo {author} {\bibfnamefont {S.~A.~R.}\ \bibnamefont {Ellis}}, \bibinfo {author} {\bibfnamefont {R.}~\bibnamefont {Harnik}}, \bibinfo {author} {\bibfnamefont {Y.}~\bibnamefont {Kahn}},\ and\ \bibinfo {author} {\bibfnamefont {J.}~\bibnamefont {Sch\"utte-Engel}},\ }\bibfield  {title} {\bibinfo {title} {{Detecting high-frequency gravitational waves with microwave cavities}},\ }\href {https://doi.org/10.1103/PhysRevD.105.116011} {\bibfield  {journal} {\bibinfo  {journal} {Phys. Rev. D}\ }\textbf {\bibinfo {volume} {105}},\ \bibinfo {pages} {116011} (\bibinfo {year} {2022})},\ \Eprint {https://arxiv.org/abs/2112.11465} {arXiv:2112.11465 [hep-ph]} \BibitemShut {NoStop}%
\bibitem [{\citenamefont {Berlin}\ \emph {et~al.}(2023)\citenamefont {Berlin}, \citenamefont {Blas}, \citenamefont {Tito~D'Agnolo}, \citenamefont {Ellis}, \citenamefont {Harnik}, \citenamefont {Kahn}, \citenamefont {Sch\"utte-Engel},\ and\ \citenamefont {Wentzel}}]{Berlin:2023grv}%
  \BibitemOpen
  \bibfield  {author} {\bibinfo {author} {\bibfnamefont {A.}~\bibnamefont {Berlin}}, \bibinfo {author} {\bibfnamefont {D.}~\bibnamefont {Blas}}, \bibinfo {author} {\bibfnamefont {R.}~\bibnamefont {Tito~D'Agnolo}}, \bibinfo {author} {\bibfnamefont {S.~A.~R.}\ \bibnamefont {Ellis}}, \bibinfo {author} {\bibfnamefont {R.}~\bibnamefont {Harnik}}, \bibinfo {author} {\bibfnamefont {Y.}~\bibnamefont {Kahn}}, \bibinfo {author} {\bibfnamefont {J.}~\bibnamefont {Sch\"utte-Engel}},\ and\ \bibinfo {author} {\bibfnamefont {M.}~\bibnamefont {Wentzel}},\ }\bibfield  {title} {\bibinfo {title} {{Electromagnetic cavities as mechanical bars for gravitational waves}},\ }\href {https://doi.org/10.1103/PhysRevD.108.084058} {\bibfield  {journal} {\bibinfo  {journal} {Phys. Rev. D}\ }\textbf {\bibinfo {volume} {108}},\ \bibinfo {pages} {084058} (\bibinfo {year} {2023})},\ \Eprint {https://arxiv.org/abs/2303.01518} {arXiv:2303.01518 [hep-ph]} \BibitemShut {NoStop}%
\bibitem [{\citenamefont {Ratzinger}\ \emph {et~al.}(2024)\citenamefont {Ratzinger}, \citenamefont {Schenk},\ and\ \citenamefont {Schwaller}}]{Ratzinger:2024spd}%
  \BibitemOpen
  \bibfield  {author} {\bibinfo {author} {\bibfnamefont {W.}~\bibnamefont {Ratzinger}}, \bibinfo {author} {\bibfnamefont {S.}~\bibnamefont {Schenk}},\ and\ \bibinfo {author} {\bibfnamefont {P.}~\bibnamefont {Schwaller}},\ }\href@noop {} {\bibinfo {title} {{A Coordinate-Independent Formalism for Detecting High-Frequency Gravitational Waves}}} (\bibinfo {year} {2024}),\ \Eprint {https://arxiv.org/abs/2404.08572} {arXiv:2404.08572 [gr-qc]} \BibitemShut {NoStop}%
\bibitem [{\citenamefont {Armengaud}\ \emph {et~al.}(2014)\citenamefont {Armengaud} \emph {et~al.}}]{IAXO2014}%
  \BibitemOpen
  \bibfield  {author} {\bibinfo {author} {\bibfnamefont {E.}~\bibnamefont {Armengaud}} \emph {et~al.},\ }\bibfield  {title} {\bibinfo {title} {{Conceptual Design of the International Axion Observatory (IAXO)}},\ }\href {https://doi.org/10.1088/1748-0221/9/05/T05002} {\bibfield  {journal} {\bibinfo  {journal} {JINST}\ }\textbf {\bibinfo {volume} {9}},\ \bibinfo {pages} {T05002}},\ \Eprint {https://arxiv.org/abs/1401.3233} {arXiv:1401.3233 [physics.ins-det]} \BibitemShut {NoStop}%
\bibitem [{\citenamefont {Armengaud}\ \emph {et~al.}(2019)\citenamefont {Armengaud} \emph {et~al.}}]{IAXO2019}%
  \BibitemOpen
  \bibfield  {author} {\bibinfo {author} {\bibfnamefont {E.}~\bibnamefont {Armengaud}} \emph {et~al.} (\bibinfo {collaboration} {IAXO}),\ }\bibfield  {title} {\bibinfo {title} {{Physics potential of the {I}nternational {A}xion {O}bservatory ({IAXO})}},\ }\href {https://doi.org/10.1088/1475-7516/2019/06/047} {\bibfield  {journal} {\bibinfo  {journal} {JCAP}\ }\textbf {\bibinfo {volume} {06}},\ \bibinfo {pages} {047}},\ \Eprint {https://arxiv.org/abs/1904.09155} {arXiv:1904.09155 [hep-ph]} \BibitemShut {NoStop}%
\bibitem [{\citenamefont {Abeln}\ \emph {et~al.}(2021)\citenamefont {Abeln}, \citenamefont {Altenmüller}, \citenamefont {Cuendis}, \citenamefont {Armengaud}, \citenamefont {Attié}, \citenamefont {Aune}, \citenamefont {Basso}, \citenamefont {Bergé}, \citenamefont {Biasuzzi}, \citenamefont {Borges~de Sousa}, \citenamefont {Brun}, \citenamefont {Bykovskiy}, \citenamefont {Calvet}, \citenamefont {Carmona}, \citenamefont {Castel}, \citenamefont {Cebrián}, \citenamefont {Chernov}, \citenamefont {Christensen}, \citenamefont {Civitani},\ and\ \citenamefont {Yanes-Díaz}}]{BabyIAXO2021}%
  \BibitemOpen
  \bibfield  {author} {\bibinfo {author} {\bibfnamefont {A.}~\bibnamefont {Abeln}}, \bibinfo {author} {\bibfnamefont {K.}~\bibnamefont {Altenmüller}}, \bibinfo {author} {\bibfnamefont {S.}~\bibnamefont {Cuendis}}, \bibinfo {author} {\bibfnamefont {E.}~\bibnamefont {Armengaud}}, \bibinfo {author} {\bibfnamefont {D.}~\bibnamefont {Attié}}, \bibinfo {author} {\bibfnamefont {S.}~\bibnamefont {Aune}}, \bibinfo {author} {\bibfnamefont {S.}~\bibnamefont {Basso}}, \bibinfo {author} {\bibfnamefont {L.}~\bibnamefont {Bergé}}, \bibinfo {author} {\bibfnamefont {B.}~\bibnamefont {Biasuzzi}}, \bibinfo {author} {\bibfnamefont {P.}~\bibnamefont {Borges~de Sousa}}, \bibinfo {author} {\bibfnamefont {P.}~\bibnamefont {Brun}}, \bibinfo {author} {\bibfnamefont {N.}~\bibnamefont {Bykovskiy}}, \bibinfo {author} {\bibfnamefont {D.}~\bibnamefont {Calvet}}, \bibinfo {author} {\bibfnamefont {J.}~\bibnamefont {Carmona}}, \bibinfo {author} {\bibfnamefont {J.}~\bibnamefont {Castel}}, \bibinfo {author} {\bibfnamefont {S.}~\bibnamefont
  {Cebrián}}, \bibinfo {author} {\bibfnamefont {V.}~\bibnamefont {Chernov}}, \bibinfo {author} {\bibfnamefont {F.}~\bibnamefont {Christensen}}, \bibinfo {author} {\bibfnamefont {M.}~\bibnamefont {Civitani}},\ and\ \bibinfo {author} {\bibfnamefont {A.}~\bibnamefont {Yanes-Díaz}},\ }\bibfield  {title} {\bibinfo {title} {Conceptual design of {B}aby{IAXO}, the intermediate stage towards the {I}nternational {A}xion {O}bservatory},\ }\href {https://doi.org/10.1007/JHEP05(2021)137} {\bibfield  {journal} {\bibinfo  {journal} {Journal of High Energy Physics}\ }\textbf {\bibinfo {volume} {2021}} (\bibinfo {year} {2021})}\BibitemShut {NoStop}%
\bibitem [{\citenamefont {Ahyoune}\ \emph {et~al.}(2023)\citenamefont {Ahyoune} \emph {et~al.}}]{BABYIAXO_RADES}%
  \BibitemOpen
  \bibfield  {author} {\bibinfo {author} {\bibfnamefont {S.}~\bibnamefont {Ahyoune}} \emph {et~al.},\ }\bibfield  {title} {\bibinfo {title} {{A {P}roposal for a {L}ow-{F}requency {A}xion {S}earch in the 1-2 $\mu$e{V} {R}ange and {B}elow with the {B}aby{IAXO} {M}agnet}},\ }\href {https://doi.org/10.1002/andp.202300326} {\bibfield  {journal} {\bibinfo  {journal} {Annalen Phys.}\ }\textbf {\bibinfo {volume} {535}},\ \bibinfo {pages} {2300326} (\bibinfo {year} {2023})},\ \Eprint {https://arxiv.org/abs/2306.17243} {arXiv:2306.17243 [physics.ins-det]} \BibitemShut {NoStop}%
\bibitem [{\citenamefont {Siodlaczek}(2022)}]{cryostat}%
  \BibitemOpen
  \bibfield  {author} {\bibinfo {author} {\bibfnamefont {M.}~\bibnamefont {Siodlaczek}},\ }\emph {\bibinfo {title} {{Thermodynamic Design of a Cryostat for a Radiofrequency Cavity Detector in BabyIAXO in Search of Dark Matter}}},\ \href@noop {} {Master's thesis},\ \bibinfo  {school} {CERN} (\bibinfo {year} {2022})\BibitemShut {NoStop}%
\bibitem [{\citenamefont {Pozar}(2005)}]{Pozar}%
  \BibitemOpen
  \bibfield  {author} {\bibinfo {author} {\bibfnamefont {D.~M.}\ \bibnamefont {Pozar}},\ }\href@noop {} {\emph {\bibinfo {title} {{Microwave engineering; 3rd ed.}}}}\ (\bibinfo  {publisher} {Wiley},\ \bibinfo {address} {Hoboken, NJ},\ \bibinfo {year} {2005})\BibitemShut {NoStop}%
\bibitem [{\citenamefont {Navarro}\ \emph {et~al.}(2024)\citenamefont {Navarro}, \citenamefont {Gimeno}, \citenamefont {Monzó-Cabrera}, \citenamefont {Diaz-Morcillo},\ and\ \citenamefont {Blas}}]{HFGW_pablo}%
  \BibitemOpen
  \bibfield  {author} {\bibinfo {author} {\bibfnamefont {P.}~\bibnamefont {Navarro}}, \bibinfo {author} {\bibfnamefont {B.}~\bibnamefont {Gimeno}}, \bibinfo {author} {\bibfnamefont {J.}~\bibnamefont {Monzó-Cabrera}}, \bibinfo {author} {\bibfnamefont {A.}~\bibnamefont {Diaz-Morcillo}},\ and\ \bibinfo {author} {\bibfnamefont {D.}~\bibnamefont {Blas}},\ }\bibfield  {title} {\bibinfo {title} {Study of a cubic cavity resonator for gravitational waves detection in the microwave frequency range},\ }\href {https://doi.org/10.1103/PhysRevD.109.104048} {\bibfield  {journal} {\bibinfo  {journal} {Physical Review D}\ }\textbf {\bibinfo {volume} {109}} (\bibinfo {year} {2024})}\BibitemShut {NoStop}%
\bibitem [{\citenamefont {Domcke}\ \emph {et~al.}(2022)\citenamefont {Domcke}, \citenamefont {Garcia-Cely},\ and\ \citenamefont {Rodd}}]{PhysRevLett.129.041101}%
  \BibitemOpen
  \bibfield  {author} {\bibinfo {author} {\bibfnamefont {V.}~\bibnamefont {Domcke}}, \bibinfo {author} {\bibfnamefont {C.}~\bibnamefont {Garcia-Cely}},\ and\ \bibinfo {author} {\bibfnamefont {N.~L.}\ \bibnamefont {Rodd}},\ }\bibfield  {title} {\bibinfo {title} {Novel search for high-frequency gravitational waves with low-mass axion haloscopes},\ }\href {https://doi.org/10.1103/PhysRevLett.129.041101} {\bibfield  {journal} {\bibinfo  {journal} {Phys. Rev. Lett.}\ }\textbf {\bibinfo {volume} {129}},\ \bibinfo {pages} {041101} (\bibinfo {year} {2022})}\BibitemShut {NoStop}%
\bibitem [{\citenamefont {{CST AG}}(2024)}]{CSTStudioSuite}%
  \BibitemOpen
  \bibfield  {author} {\bibinfo {author} {\bibnamefont {{CST AG}}},\ }\href {https://www.3ds.com/products-services/simulia/products/cst-studio-suite/} {\bibinfo {title} {Cst studio suite}} (\bibinfo {year} {2024}),\ \bibinfo {note} {versión 2024}\BibitemShut {NoStop}%
\bibitem [{Note1()}]{Note1}%
  \BibitemOpen
  \bibinfo {note} {Nevertheless, as can be seen in \cite {BABYIAXO_RADES}, since it has some distortions in the external part of the bore, the magnetic field of the BabyIAXO experiment is not completely uniform. This aspect is more relevant in Cavity 1 since its dimensions are higher}\BibitemShut {NoStop}%
\bibitem [{Note2()}]{Note2}%
  \BibitemOpen
  \bibinfo {note} {Excluding the possibility of two axions with a mass split corresponding to the one between both modes.}\BibitemShut {Stop}%
\bibitem [{\citenamefont {Knirck}\ \emph {et~al.}(2018)\citenamefont {Knirck}, \citenamefont {Millar}, \citenamefont {O'Hare}, \citenamefont {Redondo},\ and\ \citenamefont {Steffen}}]{Knirck_2018}%
  \BibitemOpen
  \bibfield  {author} {\bibinfo {author} {\bibfnamefont {S.}~\bibnamefont {Knirck}}, \bibinfo {author} {\bibfnamefont {A.~J.}\ \bibnamefont {Millar}}, \bibinfo {author} {\bibfnamefont {C.~A.}\ \bibnamefont {O'Hare}}, \bibinfo {author} {\bibfnamefont {J.}~\bibnamefont {Redondo}},\ and\ \bibinfo {author} {\bibfnamefont {F.~D.}\ \bibnamefont {Steffen}},\ }\bibfield  {title} {\bibinfo {title} {Directional axion detection},\ }\href {https://doi.org/10.1088/1475-7516/2018/11/051} {\bibfield  {journal} {\bibinfo  {journal} {Journal of Cosmology and Astroparticle Physics}\ }\textbf {\bibinfo {volume} {2018}}\bibinfo  {number} { (11)},\ \bibinfo {pages} {051}}\BibitemShut {NoStop}%
\bibitem [{\citenamefont {Irastorza}\ and\ \citenamefont {García}(2012)}]{Igor_directional}%
  \BibitemOpen
\bibfield  {number} {  }\bibfield  {author} {\bibinfo {author} {\bibfnamefont {I.~G.}\ \bibnamefont {Irastorza}}\ and\ \bibinfo {author} {\bibfnamefont {J.~A.}\ \bibnamefont {García}},\ }\bibfield  {title} {\bibinfo {title} {Direct detection of dark matter axions with directional sensitivity},\ }\href {https://doi.org/10.1088/1475-7516/2012/10/022} {\bibfield  {journal} {\bibinfo  {journal} {Journal of Cosmology and Astroparticle Physics}\ }\textbf {\bibinfo {volume} {2012}}\bibinfo  {number} { (10)},\ \bibinfo {pages} {022}}\BibitemShut {NoStop}%
\bibitem [{\citenamefont {Maggiore}(2007)}]{Maggiore}%
  \BibitemOpen
\bibfield  {number} {  }\bibfield  {author} {\bibinfo {author} {\bibfnamefont {M.}~\bibnamefont {Maggiore}},\ }\href@noop {} {\emph {\bibinfo {title} {{Gravitational Waves. Vol. 1: Theory and Experiments}}}},\ Oxford Master Series in Physics\ (\bibinfo  {publisher} {Oxford University Press},\ \bibinfo {year} {2007})\BibitemShut {NoStop}%
\bibitem [{\citenamefont {Franciolini}\ \emph {et~al.}(2022)\citenamefont {Franciolini}, \citenamefont {Maharana},\ and\ \citenamefont {Muia}}]{Franciolini:2022htd}%
  \BibitemOpen
  \bibfield  {author} {\bibinfo {author} {\bibfnamefont {G.}~\bibnamefont {Franciolini}}, \bibinfo {author} {\bibfnamefont {A.}~\bibnamefont {Maharana}},\ and\ \bibinfo {author} {\bibfnamefont {F.}~\bibnamefont {Muia}},\ }\bibfield  {title} {\bibinfo {title} {{Hunt for light primordial black hole dark matter with ultrahigh-frequency gravitational waves}},\ }\href {https://doi.org/10.1103/PhysRevD.106.103520} {\bibfield  {journal} {\bibinfo  {journal} {Phys. Rev. D}\ }\textbf {\bibinfo {volume} {106}},\ \bibinfo {pages} {103520} (\bibinfo {year} {2022})},\ \Eprint {https://arxiv.org/abs/2205.02153} {arXiv:2205.02153 [astro-ph.CO]} \BibitemShut {NoStop}%
\bibitem [{\citenamefont {Brito}\ \emph {et~al.}(2015)\citenamefont {Brito}, \citenamefont {Cardoso},\ and\ \citenamefont {Pani}}]{Brito:2015oca}%
  \BibitemOpen
  \bibfield  {author} {\bibinfo {author} {\bibfnamefont {R.}~\bibnamefont {Brito}}, \bibinfo {author} {\bibfnamefont {V.}~\bibnamefont {Cardoso}},\ and\ \bibinfo {author} {\bibfnamefont {P.}~\bibnamefont {Pani}},\ }\bibfield  {title} {\bibinfo {title} {{Superradiance}: {New Frontiers in Black Hole Physics}},\ }\href {https://doi.org/10.1007/978-3-319-19000-6} {\bibfield  {journal} {\bibinfo  {journal} {Lect. Notes Phys.}\ }\textbf {\bibinfo {volume} {906}},\ \bibinfo {pages} {pp.1} (\bibinfo {year} {2015})},\ \Eprint {https://arxiv.org/abs/1501.06570} {arXiv:1501.06570 [gr-qc]} \BibitemShut {NoStop}%
\bibitem [{\citenamefont {Kim}\ \emph {et~al.}(2020)\citenamefont {Kim}, \citenamefont {Jeong}, \citenamefont {Youn}, \citenamefont {Kim},\ and\ \citenamefont {Semertzidis}}]{detection_rate_JCAP}%
  \BibitemOpen
  \bibfield  {author} {\bibinfo {author} {\bibfnamefont {D.}~\bibnamefont {Kim}}, \bibinfo {author} {\bibfnamefont {J.}~\bibnamefont {Jeong}}, \bibinfo {author} {\bibfnamefont {S.}~\bibnamefont {Youn}}, \bibinfo {author} {\bibfnamefont {Y.}~\bibnamefont {Kim}},\ and\ \bibinfo {author} {\bibfnamefont {Y.~K.}\ \bibnamefont {Semertzidis}},\ }\bibfield  {title} {\bibinfo {title} {Revisiting the detection rate for axion haloscopes},\ }\href {https://doi.org/10.1088/1475-7516/2020/03/066} {\bibfield  {journal} {\bibinfo  {journal} {Journal of Cosmology and Astroparticle Physics}\ }\textbf {\bibinfo {volume} {2020}}\bibinfo  {number} { (03)},\ \bibinfo {pages} {066}}\BibitemShut {NoStop}%
\bibitem [{\citenamefont {Golm}\ \emph {et~al.}(2022)\citenamefont {Golm}, \citenamefont {Arguedas~Cuendis}, \citenamefont {Calatroni}, \citenamefont {Cogollos}, \citenamefont {Döbrich}, \citenamefont {Gallego}, \citenamefont {García~Barceló}, \citenamefont {Granados}, \citenamefont {Gutierrez}, \citenamefont {Irastorza}, \citenamefont {Koettig}, \citenamefont {Lamas}, \citenamefont {Liberadzka-Porret}, \citenamefont {Malbrunot}, \citenamefont {Millar}, \citenamefont {Navarro}, \citenamefont {Carlos}, \citenamefont {Puig}, \citenamefont {Rosaz}, \citenamefont {Siodlaczek}, \citenamefont {Telles},\ and\ \citenamefont {Wuensch}}]{superconducting_cavities}%
  \BibitemOpen
\bibfield  {number} {  }\bibfield  {author} {\bibinfo {author} {\bibfnamefont {J.}~\bibnamefont {Golm}}, \bibinfo {author} {\bibfnamefont {S.}~\bibnamefont {Arguedas~Cuendis}}, \bibinfo {author} {\bibfnamefont {S.}~\bibnamefont {Calatroni}}, \bibinfo {author} {\bibfnamefont {C.}~\bibnamefont {Cogollos}}, \bibinfo {author} {\bibfnamefont {B.}~\bibnamefont {Döbrich}}, \bibinfo {author} {\bibfnamefont {J.}~\bibnamefont {Gallego}}, \bibinfo {author} {\bibfnamefont {J.}~\bibnamefont {García~Barceló}}, \bibinfo {author} {\bibfnamefont {X.}~\bibnamefont {Granados}}, \bibinfo {author} {\bibfnamefont {J.}~\bibnamefont {Gutierrez}}, \bibinfo {author} {\bibfnamefont {I.}~\bibnamefont {Irastorza}}, \bibinfo {author} {\bibfnamefont {T.}~\bibnamefont {Koettig}}, \bibinfo {author} {\bibfnamefont {N.}~\bibnamefont {Lamas}}, \bibinfo {author} {\bibfnamefont {J.}~\bibnamefont {Liberadzka-Porret}}, \bibinfo {author} {\bibfnamefont {C.}~\bibnamefont {Malbrunot}}, \bibinfo {author} {\bibfnamefont {W.}~\bibnamefont {Millar}},
  \bibinfo {author} {\bibfnamefont {P.}~\bibnamefont {Navarro}}, \bibinfo {author} {\bibfnamefont {C.}~\bibnamefont {Carlos}}, \bibinfo {author} {\bibfnamefont {T.}~\bibnamefont {Puig}}, \bibinfo {author} {\bibfnamefont {G.}~\bibnamefont {Rosaz}}, \bibinfo {author} {\bibfnamefont {M.}~\bibnamefont {Siodlaczek}}, \bibinfo {author} {\bibfnamefont {G.}~\bibnamefont {Telles}},\ and\ \bibinfo {author} {\bibfnamefont {W.}~\bibnamefont {Wuensch}},\ }\bibfield  {title} {\bibinfo {title} {Thin film (high temperature) superconducting radiofrequency cavities for the search of axion dark matter},\ }\href {https://doi.org/10.1109/TASC.2022.3147741} {\bibfield  {journal} {\bibinfo  {journal} {IEEE Transactions on Applied Superconductivity}\ }\textbf {\bibinfo {volume} {32}},\ \bibinfo {pages} {1} (\bibinfo {year} {2022})}\BibitemShut {NoStop}%
\bibitem [{\citenamefont {Gatti}\ \emph {et~al.}(2024)\citenamefont {Gatti}, \citenamefont {Visinelli},\ and\ \citenamefont {Zantedeschi}}]{FLASH}%
  \BibitemOpen
  \bibfield  {author} {\bibinfo {author} {\bibfnamefont {C.}~\bibnamefont {Gatti}}, \bibinfo {author} {\bibfnamefont {L.}~\bibnamefont {Visinelli}},\ and\ \bibinfo {author} {\bibfnamefont {M.}~\bibnamefont {Zantedeschi}},\ }\bibfield  {title} {\bibinfo {title} {{Cavity detection of gravitational waves: Where do we stand?}},\ }\href {https://doi.org/10.1103/PhysRevD.110.023018} {\bibfield  {journal} {\bibinfo  {journal} {Phys. Rev. D}\ }\textbf {\bibinfo {volume} {110}},\ \bibinfo {pages} {023018} (\bibinfo {year} {2024})},\ \Eprint {https://arxiv.org/abs/2403.18610} {arXiv:2403.18610 [gr-qc]} \BibitemShut {NoStop}%
\bibitem [{\citenamefont {Díaz-Morcillo}\ \emph {et~al.}(2022)\citenamefont {Díaz-Morcillo}, \citenamefont {García~Barceló}, \citenamefont {Lozano~Guerrero}, \citenamefont {Navarro}, \citenamefont {Gimeno}, \citenamefont {Arguedas~Cuendis}, \citenamefont {Álvarez Melcón}, \citenamefont {Cogollos}, \citenamefont {Calatroni}, \citenamefont {Döbrich}, \citenamefont {Gallego-Puyol}, \citenamefont {Golm}, \citenamefont {Irastorza}, \citenamefont {Malbrunot}, \citenamefont {Miralda-Escudé}, \citenamefont {Peña~Garay}, \citenamefont {Redondo},\ and\ \citenamefont {Wuensch}}]{universe_RADES}%
  \BibitemOpen
  \bibfield  {author} {\bibinfo {author} {\bibfnamefont {A.}~\bibnamefont {Díaz-Morcillo}}, \bibinfo {author} {\bibfnamefont {J.~M.}\ \bibnamefont {García~Barceló}}, \bibinfo {author} {\bibfnamefont {A.~J.}\ \bibnamefont {Lozano~Guerrero}}, \bibinfo {author} {\bibfnamefont {P.}~\bibnamefont {Navarro}}, \bibinfo {author} {\bibfnamefont {B.}~\bibnamefont {Gimeno}}, \bibinfo {author} {\bibfnamefont {S.}~\bibnamefont {Arguedas~Cuendis}}, \bibinfo {author} {\bibfnamefont {A.}~\bibnamefont {Álvarez Melcón}}, \bibinfo {author} {\bibfnamefont {C.}~\bibnamefont {Cogollos}}, \bibinfo {author} {\bibfnamefont {S.}~\bibnamefont {Calatroni}}, \bibinfo {author} {\bibfnamefont {B.}~\bibnamefont {Döbrich}}, \bibinfo {author} {\bibfnamefont {J.~D.}\ \bibnamefont {Gallego-Puyol}}, \bibinfo {author} {\bibfnamefont {J.}~\bibnamefont {Golm}}, \bibinfo {author} {\bibfnamefont {I.~G.}\ \bibnamefont {Irastorza}}, \bibinfo {author} {\bibfnamefont {C.}~\bibnamefont {Malbrunot}}, \bibinfo {author} {\bibfnamefont {J.}~\bibnamefont
  {Miralda-Escudé}}, \bibinfo {author} {\bibfnamefont {C.}~\bibnamefont {Peña~Garay}}, \bibinfo {author} {\bibfnamefont {J.}~\bibnamefont {Redondo}},\ and\ \bibinfo {author} {\bibfnamefont {W.}~\bibnamefont {Wuensch}},\ }\bibfield  {title} {\bibinfo {title} {Design of new resonant haloscopes in the search for the dark matter axion: A review of the first steps in the rades collaboration},\ }\bibfield  {journal} {\bibinfo  {journal} {Universe}\ }\textbf {\bibinfo {volume} {8}},\ \href {https://doi.org/10.3390/universe8010005} {10.3390/universe8010005} (\bibinfo {year} {2022})\BibitemShut {NoStop}%
\bibitem [{\citenamefont {Thomson}\ \emph {et~al.}(2023)\citenamefont {Thomson}, \citenamefont {Goryachev}, \citenamefont {McAllister}, \citenamefont {Ivanov}, \citenamefont {Altin},\ and\ \citenamefont {Tobar}}]{australianos}%
  \BibitemOpen
  \bibfield  {author} {\bibinfo {author} {\bibfnamefont {C.~A.}\ \bibnamefont {Thomson}}, \bibinfo {author} {\bibfnamefont {M.}~\bibnamefont {Goryachev}}, \bibinfo {author} {\bibfnamefont {B.~T.}\ \bibnamefont {McAllister}}, \bibinfo {author} {\bibfnamefont {E.~N.}\ \bibnamefont {Ivanov}}, \bibinfo {author} {\bibfnamefont {P.}~\bibnamefont {Altin}},\ and\ \bibinfo {author} {\bibfnamefont {M.~E.}\ \bibnamefont {Tobar}},\ }\bibfield  {title} {\bibinfo {title} {Searching for low-mass axions using resonant upconversion},\ }\href {https://doi.org/10.1103/PhysRevD.107.112003} {\bibfield  {journal} {\bibinfo  {journal} {Phys. Rev. D}\ }\textbf {\bibinfo {volume} {107}},\ \bibinfo {pages} {112003} (\bibinfo {year} {2023})}\BibitemShut {NoStop}%
\bibitem [{Note3()}]{Note3}%
  \BibitemOpen
  \bibinfo {note} {This is seen in Fig.~\ref {fig:freq_tuning}. However, via a optimization of the cavity dimensions this frequency gap can be set to any desired value.}\BibitemShut {Stop}%
\end{thebibliography}%

\end{document}